\documentclass[lettersize,journal]{IEEEtran}
\usepackage{amsmath,amsfonts}
\usepackage{amssymb}
\usepackage{algorithm}
\usepackage{algorithmicx}
\usepackage{array}
\usepackage[caption=false,font=normalsize,labelfont=sf,textfont=sf]{subfig}
\usepackage{textcomp}
\usepackage{stfloats}
\usepackage{url}
\usepackage{verbatim}
\usepackage{graphicx}
\usepackage{cite}
\usepackage{ulem}
\usepackage{threeparttable}
\usepackage{bm}

\usepackage[
pdfauthor={derajan},
pdftitle={How to do this},
pdfstartview=XYZ,
bookmarks=true,
colorlinks=true,
linkcolor=blue,
urlcolor=blue,
citecolor=blue,
pdftex,
bookmarks=true,
linktocpage=true,   
hyperindex=true
]{hyperref}

\usepackage{newtxmath} 
\usepackage{enumitem}  

\begin{document}

\title{PLA for Drone RID Frames via Motion Estimation and Consistency Verification}
\author{Jie Li,
	Jing Li,~\IEEEmembership{Member,~IEEE,}
	Lu Lv,~\IEEEmembership{Member,~IEEE,}
	Zhanyu Ju,
	Fengkui Gong,~\IEEEmembership{Member,~IEEE,}
	\thanks{ \textit{(Corresponding author: Jing Li.)}}
	\thanks{Jie Li, Jing Li, Zhanyu Ju, and Fengkui Gong are with the State Key Laboratory of Integrated Services Network, Xidian University, Xi'an, Shaanxi 710071, China (e-mail: lijie\_372@stu.xidian.edu.cn; jli@xidian.edu.cn; juzhanyu@stu.xidian.edu.cn; fkgong@xidian.edu.cn).}
	\thanks{Lu Lv is with the School of Telecommunications Engineering, Xidian University, Xi'an 710071, China, (e-mail: lulv@xidian.edu.cn).}
}
\maketitle
\begin{abstract}
	Drone Remote Identification (RID) plays a critical role in low-altitude airspace supervision, yet its broadcast nature and lack of cryptographic protection make it vulnerable to spoofing and replay attacks. In this paper, we propose a consistency verification-based physical-layer authentication (PLA) algorithm for drone RID frames. A RID-aware sensing and decoding module is first developed to extract communication-derived sensing parameters, including angle-of-arrival, Doppler shift, average channel gain, and the number of transmit antennas, together with the identity and motion-related information decoded from previously authenticated RID frames. Rather than fusing all heterogeneous information into a single representation, different types of information are selectively utilized according to their physical relevance and reliability. Specifically, real-time wireless sensing parameter constraints and previously authenticated motion states are incorporated in a yaw-augmented constant-acceleration extended Kalman filter (CA-EKF) to estimate the three-dimensional position and motion states of the drone. To further enhance authentication reliability under highly maneuverable and non-stationary flight scenarios, a data-driven long short-term memory-based motion estimator is employed, and its predictions are adaptively combined with the CA-EKF via an error-aware fusion strategy. Finally, RID frames are authenticated by verifying consistency in the number of transmit antennas, motion estimates, and no-fly-zone constraints. Simulation results demonstrate that the proposed algorithm significantly improves authentication reliability and robustness under realistic wireless impairments and complex drone maneuvers, outperforming existing RF feature-based and motion model-based PLA schemes.
\end{abstract}


\section{Introduction}
\label{sec1}
\IEEEPARstart{T}{he} rapid proliferation of unmanned aerial vehicles, also known as drones, has enabled a wide range of civilian and commercial applications, including aerial sensing, logistics, infrastructure inspection, and emergency response. Meanwhile, the increasing density of drone operations in low-altitude airspace has raised significant concerns regarding airspace safety, spectrum coexistence, and regulatory compliance. To address these challenges, aviation authorities worldwide have introduced Remote Identification (RID) regulations, which require drones to periodically broadcast identification and motion-related information for traffic monitoring and accountability\textcolor{blue}{\cite{ref_syn}}. Conceptually, drone RID plays a role analogous to the Automatic Dependent Surveillance-Broadcast (ADS-B) system in manned aviation\textcolor{blue}{\cite{ref_adsb}}, serving as a foundational mechanism for cooperative airspace awareness.

Despite its regulatory importance, RID frames are commonly broadcast in plaintext without encryption or cryptographic authentication over open wireless channels, rendering them vulnerable to spoofing, replay, and impersonation attacks\textcolor{blue}{\cite{ref_survey1}}. Conventional cryptographic authentication schemes typically incur non-negligible computational and memory overhead, as well as additional latency due to the generation and verification of authentication tags, which may not be well suited for resource-constrained drone platforms\textcolor{blue}{\cite{ref_survey2}}. These limitations motivate the adoption of physical-layer authentication (PLA) techniques, which exploit inherent wireless and motion-related characteristics to verify the legitimacy of signals\textcolor{blue}{\cite{ref_pla}}. 

However, most existing PLA methods rely on either instantaneous radio-frequency (RF) features or idealized motion models, while overlooking the heterogeneous information available in drone RID frames. In practice, RID frames provide a rich set of information, including physical-layer parameters estimated from the received signals (e.g., angles, Doppler shifts, and channel gains), as well as identity and motion-related parameters directly decoded from RID frames. Moreover, authentication methods based solely on simplified motion models are often inadequate for capturing the highly maneuverable and non-stationary motion behaviors of drones. Therefore, how to effectively incorporate heterogeneous RID-based sensing and decoding information into communication-aware motion state estimation, and further integrate it with data-driven estimators to achieve robust consistency verification-based PLA, remains a largely unexplored challenge.

\subsection{Related Works}
\begin{enumerate}[leftmargin=0pt, itemindent=2pc, listparindent=\parindent]
	\item{ \textit{Cryptography-based Authentication}: Cryptography-based authentication has been widely studied as a potential solution for RID or ADS-B signal authentication, relying on digital signatures, hash-based message authentication codes, or public key infrastructures to ensure message integrity and source authenticity\textcolor{blue}{\cite{ref_survey3}}. In asymmetric key cryptography, the RID message was hashed using the Secure Hash Algorithm-256 and signed by elliptic curve cryptography with the transmitter's private key, then the receiver will verify the authenticity of the RID message using the transmitter's private key extracted from the attached X.509 certificate\textcolor{blue}{\cite{ref_hash}},\textcolor{blue}{\cite{ref_huiyi}}. Three protocols were proposed to help drones maintain anonymity while broadcasting RID frames based on digital signatures, where the universally unique identifier‌ (UUID) can only be verified through interaction with a trusted third party\textcolor{blue}{\cite{ref_alg3}}.
		
	\begin{table*}[!t]   
		\begin{center}   
			\caption{Comparison of This Work with Existing Works}  
			\label{tab:table_ref} 
			\begin{threeparttable}
				\begin{tabular}{|c|c|c|}   
					\hline   \textbf{Ref.} & \textbf{RF Features} & \textbf{Motion Models} \\   
					\hline   \textcolor{blue}{\cite{ref_268}} & RSS & \\ 
					\hline   \textcolor{blue}{\cite{ref_372}} & SNR difference & \\  
					\hline   \textcolor{blue}{\cite{ref_269}} & Doppler shift & \\
					\hline   \textcolor{blue}{\cite{ref_276}} & Angle delay power spectrum & \\
					\hline   \textcolor{blue}{\cite{ref_301}} & Channel frequency response and power delay profile & \\ 
					\hline   \textcolor{blue}{\cite{ref_310}} & Preamble with CFO & \\ 
					\hline   \textcolor{blue}{\cite{ref_339}} & AoA and channel gain & \\ 
					\hline   \textcolor{blue}{\cite{ref_343}} & Estimated distance, angle, radial velocity, and angular velocities & UKF-based CV, CA, and CTRV \\
					\hline   \textcolor{blue}{\cite{ref_307}} & ADS-B trajectory & LSTM network \\
					\hline   \textcolor{blue}{\cite{ref_334}} & TDoA & EKF-based CV, CA, and CTRV \\
					\hline   This work & AoA, ACG, Doppler shift, the number of transmit antennas, drone type, and motion parameters & yaw-augmented CA-EKF and LSTM \\
					\hline   
				\end{tabular}
			\end{threeparttable}
		\end{center}   
	\end{table*}
	
	While cryptography-based authentication can provide strong upper-layer security guarantees, its application to ADS-B or RID signals faces inherent limitations. Introducing encryption mechanisms often requires protocol modifications, which hinders backward compatibility and international standardization\textcolor{blue}{\cite{ref_survey2}}. Moreover, cryptographic tags or signatures introduce additional authentication overhead and bandwidth consumption, conflicting with the lightweight nature of RID transmissions\textcolor{blue}{\cite{ref_survey1}}. More importantly, traditional cryptographic schemes violate the openness principle of broadcast identification, where information is intended to be universally decodable without prior key establishment. These limitations motivate the exploration of lightweight and infrastructure-free PLA methods.
	}

	\item{ \textit{PLA Based on Instantaneous RF Features}: RF-based methods exploit unique and difficult-to-forge physical-layer characteristics of wireless signals to distinguish legitimate transmitters from adversaries. Since the received signal strength (RSS) inherently reflects both channel characteristics and the geographical locations of different transmitters, RSS measurements obtained by the sensors were exploited to estimate the drone’s position. The estimated position was then compared with a predetermined flight position, and the drone was considered authenticated if the discrepancy fell within a predefined error bound\textcolor{blue}{\cite{ref_268}}. Channel gain, as another RF feature of the wireless channel, was estimated using no fewer than 50 phase-shift keying pilot symbols to enable drone communication signal authentication\textcolor{blue}{\cite{ref_270}}. Besides, RSS variations resulting from the relative motion of the drone or the receiver, reflected in power differences\textcolor{blue}{\cite{ref_160}}, signal-to-noise ratio (SNR) differences\textcolor{blue}{\cite{ref_372}}, and SNR ratios\textcolor{blue}{\cite{ref_159}} between adjacent time slots, together with motion-induced Doppler frequency shift variations\textcolor{blue}{\cite{ref_269}}, have been exploited for drone signal authentication.
	
	Channel phase-based PLA methods were shown to achieve better authentication performance than channel amplitude-based counterparts, since the channel amplitude response tends to vary significantly in high-mobility communication systems\textcolor{blue}{\cite{ref_300}}. By employing an authentication request-inquiry-response procedure to eliminate the dominant phase offset introduced by secret keys in legitimate signals, the magnitudes of the resulting channel coefficients were shown to be significantly larger compared with those of illegitimate users\textcolor{blue}{\cite{ref_304}},\textcolor{blue}{\cite{ref_311}}. Furthermore, a variety of additional channel-related features have been exploited for signal authentication, including time-frequency representations\textcolor{blue}{\cite{ref_296}}, carrier frequency offset (CFO) combined with phase noise\textcolor{blue}{\cite{ref_176}}, the angle delay power spectrum\textcolor{blue}{\cite{ref_276}}, channel frequency response together with the power delay profile\textcolor{blue}{\cite{ref_301}}, time delays\textcolor{blue}{\cite{ref_373}}, and quantization errors induced by hardware imperfections\textcolor{blue}{\cite{ref_173}}. However, many of the aforementioned works relied on the assumption that the channel state information was known, and the decoding messages could be obtained under perfect synchronization conditions. To relax these assumptions, embedded preambles such as Zadoff-Chu (ZC) sequences in signals, were exploited for signal synchronization and channel equalization\textcolor{blue}{\cite{ref_zc}}, which not only enhanced the robustness of RF features against channel impairments but were also directly leveraged for signal authentication\textcolor{blue}{\cite{ref_303}},\textcolor{blue}{\cite{ref_310}}.
	
	In addition, some PLA methods exploited location-based spatial propagation domain information, including time of flight\textcolor{blue}{\cite{ref_150}}, time of arrival\textcolor{blue}{\cite{ref_174}}, angle of arrival (AoA)\textcolor{blue}{\cite{ref_339}}, time sum of arrival, time difference of arrival (TDoA)\textcolor{blue}{\cite{ref_298}}, integrated sensing and communication waveforms based on orthogonal frequency division multiplexing (OFDM) and orthogonal time frequency space\textcolor{blue}{\cite{ref_343}},\textcolor{blue}{\cite{ref_157}}, as well as mutual coupling effects in massive antenna arrays\textcolor{blue}{\cite{ref_302}}.
	
	Despite the rich diversity of instantaneous RF features explored in the above studies, their reliability is often compromised in highly dynamic drone scenarios, as channel variations, synchronization inaccuracies, and sensing noise can cause feature fluctuations and lead to inconsistent authentication results. Moreover, instantaneous signal features cannot capture the continuous motion behavior or kinematic constraints of drone trajectories over time. As a result, recent studies have increasingly shifted toward motion model-based trajectory authentication, which exploit the temporal evolution of motion states and physically kinematic models to achieve more stable and robust authentication, particularly under time-varying wireless conditions and complex drone maneuvers.
	}
	
	\item{ \textit{Motion Model-Based Trajectory Authentication}: Since the flight direction of civil aircraft did not change arbitrarily or frequently, flight trajectories were predicted using a long short-term memory (LSTM) network by exploiting historical three-dimensional (3D) positions extracted from ADS-B frames, enabling trajectory consistency checks for authentication\textcolor{blue}{\cite{ref_307}}. In contrast, drones exhibited much higher maneuverability and were more susceptible to airflow disturbances and airframe-induced vibrations\textcolor{blue}{\cite{ref_341}},\textcolor{blue}{\cite{ref_342}}, resulting in random jitter and frequent motion pattern changes. To address such dynamics, a TDoA-based extended Kalman filter (EKF) framework was employed to evaluate estimation bias under scenarios involving frequent switching among constant-velocity (CV), constant-acceleration (CA), and constant-turn-rate-and-velocity (CTRV) motion models\textcolor{blue}{\cite{ref_334}}. Furthermore, motion parameters estimated from OFDM echoes were combined with an unscented Kalman filter (UKF) to jointly identify the drone’s motion model and predict its position in the subsequent time slot, thereby enabling trajectory-aware authentication based on location and motion continuity\textcolor{blue}{\cite{ref_343}}. Table \ref{tab:table_ref} is given for comparison this work and existing works.
	}
\end{enumerate}

\subsection{Motivations and Contributions}
Nevertheless, existing authentication methods cannot be directly applied to practical drone RID scenarios, mainly due to the following reasons:

\begin{itemize}
	\item{Insufficient exploitation of RID-specific heterogeneous information. Unlike conventional communication signals, drone RID frames are broadcast-oriented, lightweight, and strictly protocol-constrained, which fundamentally limits the applicability of many existing cryptography-based and generic RF fingerprinting methods. Most prior PLA methods are designed for generic wireless waveforms without fully exploiting the rich and heterogeneous information inherently available in RID transmissions. In particular, both physical-layer parameters estimated from the currently received RID signals and historical information extracted from previously authenticated RID frames have not been jointly leveraged in existing authentication methods.
	}
	\item{Limited robustness under complex motion dynamics and time-varying wireless conditions. Most existing PLA methods rely on ideal motion models or assume reliable and stationary sensing conditions, making them highly sensitive to channel fading, synchronization errors, and abrupt drone maneuvers. Although communication-aware parameters extracted from RID signals can provide valuable information for motion state inference and position prediction, they are rarely integrated into the motion estimation process in a systematic manner. Moreover, model-driven estimators relying solely on simplified motion model assumptions are often inadequate to capture the highly maneuverable and non-stationary motion behaviors of drones. Without incorporating data-driven representations that can learn complex and long-term motion patterns, authentication performance deteriorates significantly in realistic low-altitude environments.
	}
\end{itemize}

Motivated by the above observations, we propose a consistency verification-based PLA algorithm for drone RID frames in air-to-ground (A2G) scenarios, which jointly exploits heterogeneous RF sensing parameters and decoding information under realistic wireless impairments. The contributions of this paper are summarized as follows.

\begin{itemize}
	\item{RID-aware wireless sensing and decoding module: A RID-aware wireless sensing and decoding module is developed to jointly extract heterogeneous parameters from drone RID frames. Specifically, the spatial, power, and frequency characteristics of the received drone RID frames are exploited to estimate the AoA, average channel gain (ACG), Doppler frequency shift, and the number of transmit antennas, while the drone UUID, type, and embedded motion-related parameters are obtained through RID frame decoding. This module provides reliable and real-time inputs for subsequent motion state estimation and consistency checks.
	}
	\item{Communication-aware motion state estimator: The extracted RF sensing parameters are integrated into a yaw-augmented CA-EKF estimator consisting of two kinematic estimation stages. Previously authenticated motion states are utilized to perform the initial state prediction, while real-time sensing-derived motion information is incorporated as constraints in the second update stage to refine the predicted state. The proposed estimator enables robust tracking of drone's 3D position, velocity, and yaw under time-varying wireless sensing conditions.
	}
	\item{Error-aware multi-estimator adaptive fusion strategy: A data-driven LSTM-based motion estimator is designed to learn implicit long-term dynamic patterns from historical motion states, which is adaptively fused with the model-driven CA-EKF estimator according based on their estimated error statistics. The proposed fusion strategy enables online refinement of CA-EKF predictions and significantly enhances tracking robustness under highly maneuvering and non-stationary drone motion behaviors.
	}
	\item{Our numerical results demonstrate that the proposed algorithm achieves superior authentication reliability and robustness under A2G communication conditions. Several insights are revealed: 1) By jointly exploiting real-time wireless sensing parameters, historical decoded motion information, and multi-level consistency verification, the proposed algorithm significantly outperforms existing works, particularly in highly maneuverable flight scenarios and time-varying channels. 2) The adaptive fusion strategy effectively combines the strengths of model-driven and data-driven estimators, improving robustness against nonlinear motion dynamics. 3) The performance exhibits clear and physically interpretable dependencies on system parameters, indicating a favorable balance between estimation accuracy and practical complexity.}
\end{itemize}

\subsection{Outline and Notations}
The system model and problem formulation are presented in Section \ref{sec2}. The details and explanations of location consistency-based PLA algorithm are discussed in Section \ref{sec3}. Simulation evaluation is presented in Section \ref{sec4}. And this paper is summarized in Section \ref{sec5}. 

\(\mathbb{R}^{a\times b}\) and \(\mathbb{C}^{a\times b}\) denote the sets of \(a\times b\) real-valued and complex-valued matrices, respectively. For a complex-valued vector or matrix $\mathbf{v}$, $\|\mathbf{v}\|_2$, $\mathbf{v}^{-1}$, $\mathbf{v}^{\mathsf{T}}$, $\mathbf{v}^{*}$, and $\mathbf{v}^{\dagger}$ denote its $\ell_2$-norm, inverse, transpose, conjugate, and Hermitian transpose, respectively. $\mathbb{E}[\cdot]$ and $\mathbb{G}[\cdot]$ denote the arithmetic mean and geometric mean of a set of scalars, respectively. \(\otimes\) denotes the Hadamard product. The circularly symmetric complex Gaussian random variable \(x\) with mean \(\mu\) and variance \(\sigma^{2}\) is denoted by \(x \sim \mathcal{CN}(0, \sigma^{2})\).

\section{System Model and Problem Formulation}
\label{sec2}
\subsection{System Model}
\begin{figure*}[!t]
	\centering
	\includegraphics[width=5in]{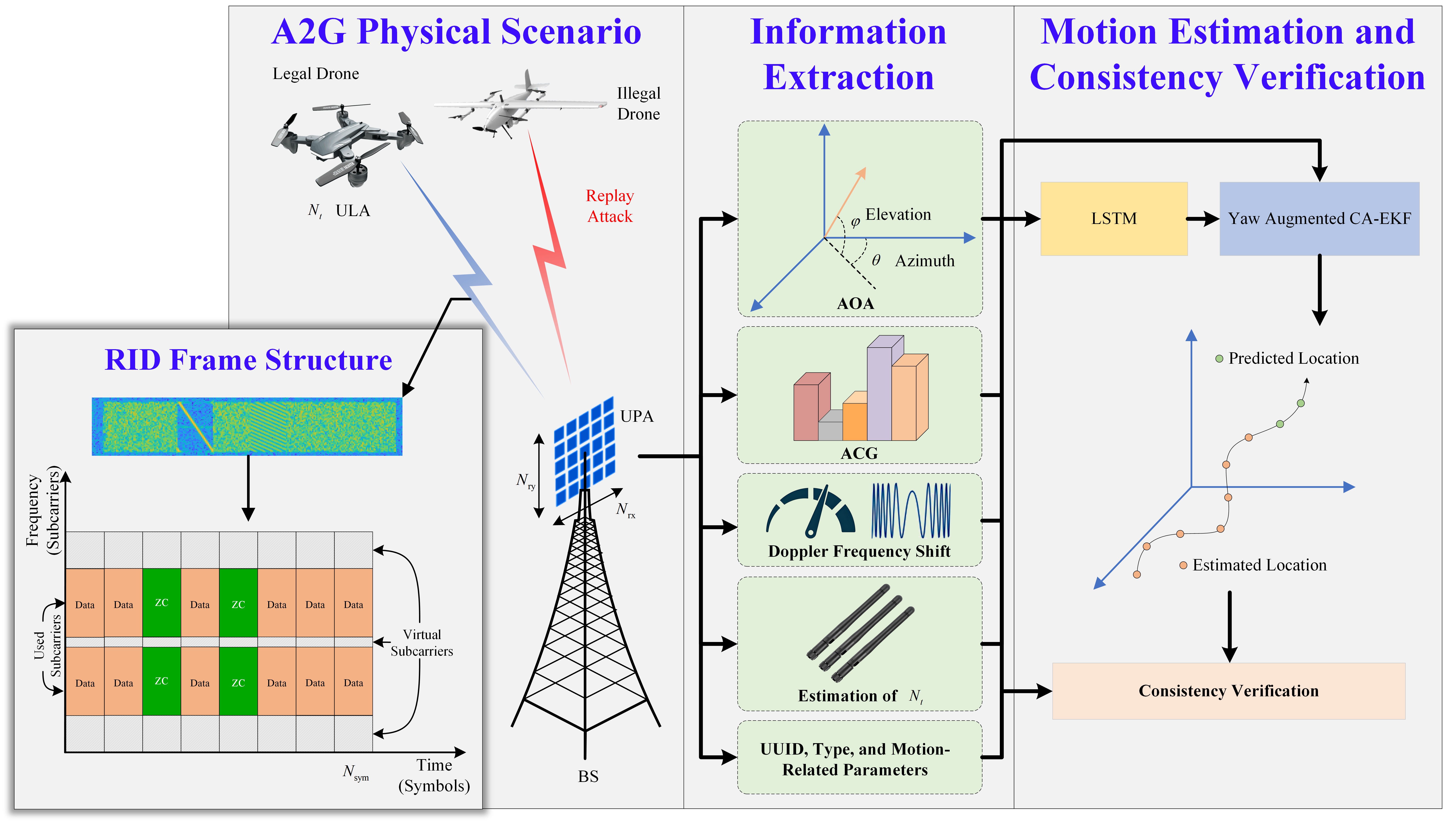}
	\caption{A sketch of consistency verification-based PLA for drone RID frames.}
	\label{Fig_Net}
\end{figure*}

We consider an A2G network as illustrated in Fig. \ref{Fig_Net}, where the base station (BS) is equipped with a uniform planar array (UPA) consisting of \(N_\text{rx} \times N_\text{ry}\) antennas, and each drone is equipped uniform linear array (ULA) with \(N_t\) antennas for transmitting RID frames. Assume that the drone flight is divided into \(T\) time slots each with a duration of \(\varpi\), the index set can be represented as \(t \in \mathcal{T} \triangleq \{1, 2, \ldots, T\}\). For \(\forall\, t \in \mathcal{T}\), the legal drone periodically broadcasts OFDM-based RID frames according to regulatory requirements, whereas an illegal drone periodically replays previously eavesdropped RID frames to deceive the BS and conceal illegal intrusion.

Specifically, an RID frame consists of \(N_\text{sym}\) OFDM symbols, including two training-sequence symbols and \(N_{\text{sym}} - 2\) data symbols, where each symbol comprises \(N\) subcarriers consisting of \(N_\text{dc}\) data subcarriers and \(N_\text{vc}\) virtual subcarriers. The training-sequence symbols are constructed from ZC sequences and are used for time-frequency synchronization and channel equalization. The data subcarriers carry modulated sequence, whereas the virtual subcarriers remain unused to suppress out-of-band emissions and mitigate adjacent-channel interference. The payload information is successively processed by Turbo coding, m-sequence-based scrambling, and quadrature phase shift keying modulation to generate the modulated sequences. Since these procedures are not the focus of this work, the underlying principles are not elaborated in detail.

Let \(\mathbf{S}_i(t)\) denote the frequency-domain RID signal frame on the \(i\)-th subcarrier transmitted by the drone at time \(t\), which can be expressed as
\begin{equation}
	\label{eq1}
	\mathbf{S}_i(t) = \bigl[ \mathbf{S}_{1,i}(t), \mathbf{S}_{2,i}(t), \ldots, \mathbf{S}_{N_{\text{sym}},i}(t) \bigr] \in \mathbb{C}^{N_t \times N_{\text{sym}}},
\end{equation}
where \(\mathbf{S}_{l,i}(t) = \bigl[ S^{1}_{l,i}(t), S^{2}_{l,i}(t), \ldots, S^{N_t}_{l,i}(t) \bigr]^{\mathsf{T}}\), and \(S^j_{l,i}(t)\) denotes the frequency-domain representation of the \(l\)-th OFDM symbol on the \(i\)-th subcarrier transmitted by antenna \(j\). The transmission frequency of \(i\)-th subcarrier can be expressed as
\begin{equation}
	\label{eqfsub}
	f_i(t) =f_c(t) + (i-i_0)\Delta f=f_c(t) + \left(i - \frac{N+1}{2}\right)\frac{B}{N},
\end{equation}
where \(f_c(t)\) and \(B\) denote the center transmission frequency and signal bandwidth, respectively. It is worth noting that a cyclic prefix (CP) of length \(N_\text{cp}\) is inserted before each symbol in the RID frame to mitigate inter-symbol interference and inter-carrier interference. The RID frame typically satisfies
\begin{equation}
	\label{eqlen}
	\frac{N + N_{\text{cp}}}{B}\, N_{\text{sym}} \ll \varpi,
\end{equation}

Then the signal received on the \(i\)-th subcarrier at the BS can be expressed as
\begin{equation}
	\label{eq2}
	\mathbf{X}_i(t)	= \bigl[ \mathbf{X}_{1,i}(t), \mathbf{X}_{2,i}(t), \ldots, \mathbf{X}_{N_{\text{sym}},i}(t) \bigr] \in \mathbb{C}^{N_r \times N_{\text{sym}}},
\end{equation}
where \(\mathbf{X}_{l,i}(t) = \bigl[ X^{1}_{l,i}(t), X^{2}_{l,i}(t), \ldots, X^{N_r}_{l,i}(t) \bigr]^{\mathsf{T}}\) and \(N_r=N_\text{rx}N_\text{ry}\). (\ref{eq2}) can be rewritten as
\begin{equation}
	\label{eq3}
	\mathbf{X}_i(t) = \mathbf{a}(\theta, \varphi)\mathbf{H}_i(t)\mathbf{S}_i(t) + \mathbf{N}_i(t),
\end{equation}
where \(\mathbf{H}_i(t) \in \mathbb{C}^{N_r \times N_t}\) and \(\mathbf{N}_i(t) \sim \mathcal{CN}\!\left( \mathbf{0}, \sigma_n^2 \mathbf{I}_{N_r} \right)\) denote the channel coefficient matrix and additive white Gaussian noise (AWGN) from the drone to the BS, respectively. The channel coefficient on the \(i\)-th subcarrier between the \(p\)-th transmit antenna and the \(q\)-th receive antenna can be expressed as\textcolor{blue}{\cite{ref_channel}}
\begin{equation}
	\label{eq4}
	\begin{aligned}
		\mathbf{H}^{i}_{q,p} = \sqrt{10^{\frac{P_r - 30}{10}}} \left( \sqrt{\frac{\mathcal{K}}{\mathcal{K} + 1}} \mathbf{h}_{i,q,p}^{\text{LoS}} + \sqrt{\frac{1}{\mathcal{K} + 1}} \mathbf{h}_{i,q,p}^{\text{NLoS}} \right),
	\end{aligned}
\end{equation}
where \(\mathcal{K}\) is Rician factors of \(3\) dB, \((\cdot)^{\text{LoS}}\) stands the deterministic line-of-sight (LoS) component, and \((\cdot)^{\text{NLoS}}\) denotes the random non-LoS component. The LoS and NLoS components incorporate the effects of multipath-induced frequency offsets and delays, as well as Doppler frequency shifts \(f_d(t)\) caused by the relative motion between the drone and the BS. \(f_d(t)\) can be computed by
\begin{equation}
	\label{eqfd}
	f_d(t) = \frac{v(t)f_c(t)\cos \theta(t)}{c},
\end{equation}
where \(v(t)\) and \(c\) denote the drone's velocity and the speed of light. \(P_r(t)\) is the power scaling factor, which can be obtained as
\begin{equation}
	\label{eqpower}
	P_r(t) = P_t(t) + G_t + G_r - P_h(t),
\end{equation}
where \(P_t(t)\), \(G_t\), \(G_r\), and \(P_h\) is the transmit power, the transmit antenna gain, the receive antenna gain, and the free-space path loss, respectively. And \(P_h(t) = 20\lg d(t) + 20\lg f_c - 147.55\), where \(d(t) = \left\| \mathbf{P}_u(t)-\mathbf{P}_b \right\|_2\) denotes the distance between the drone's 3D position \(\mathbf{P}_u(t)\) and the BS's 3D position \(\mathbf{P}_b\).

\(\mathbf{a}(\theta, \varphi)=\bigl[a_{1,1}(\theta, \varphi), \ldots, a_{1, N_{\text{ry}}}(\theta, \varphi), \ldots, a_{N_{\text{rx}}, N_{\text{ry}}}(\theta, \varphi) \bigr]\) is the spatial array response vector of the UPA, and \(a_{u,v}(\theta,\varphi)\) can be expressed as
\begin{equation}
	\label{eq5}
	a_{u,v}(\theta,\varphi) = \text{e}^{-\text{j}\frac{2\pi}{\lambda}\left(u d_x \cos \varphi \cos \theta + v d_y \cos \varphi \sin \theta \right)},
\end{equation}
where \(\theta\) is the azimuth AoA, \(\varphi\) is the elevation AoA, \(d_x=d_x=\frac{\lambda}{2}\) denotes the element spacing, and \(\lambda\) is the wavelength.
 
\subsection{Problem Formulation}
For the received drone RID signals, the number of transmitting antennas, azimuth AoA, elevation AoA, ACG, and Doppler frequency shift are estimated to infer the 3D position and velocity of the drone. Meanwhile, the drone UUID, type, and motion-related parameters are extracted by decoding previously authenticated RID frames. These heterogeneous sensing and decoding information are jointly fed into a yaw-augmented CA-EKF and a LSTM-based estimator to obtain kinematically constrained 3D position estimates. Subsequently, the estimated number of transmitting antennas is cross-validated against the claimed drone type, while the estimated position is checked against no-fly-zone (NFZ) constraints to detect abnormal or non-compliant behavior. Finally, the estimated positions \(\hat{\mathbf{P}}_u(t)\) are compared with the location information \(\mathbf{P}_u(t)\) embedded in the RID frames to perform location consistency check, thereby jointly verifying the legitimacy of the received RID frames and the authenticity of the transmitting drone.

The consistency between \(\mathbf{P}_u(t)\) and \(\hat{\mathbf{P}}_u(t)\) is evaluated via a Euclidean distance-based threshold \(\gamma\), which is learned from the neural network training set. Specifically, a binary hypothesis test is performed to determine the legitimacy of the received RID frame, which is formulated as
\begin{equation}
	\label{Eq6}
	\left\| \mathbf{P}_u(t) - \hat{\mathbf{P}}_u(t) \right\|_2 \;\underset{\mathcal{H}_1}{\overset{\mathcal{H}_0}{\lessgtr}}\; \gamma,
\end{equation}
where \(\mathcal{H}_0\) denotes that the estimated drone position is consistent with the location information broadcast in the RID frame, while \(\mathcal{H}_1\) indicates that the RID frame or the drone is illegal.

\section{Consistency Verification-based PLA Algorithm}
\label{sec3}

The proposed algorithm consists of three main components. First, a RID-aware wireless sensing and decoding module is designed to extract heterogeneous parameters from the received drone RID frames. These parameters are then fed into a communication-aware motion state estimator based on a yaw-augmented CA-EKF, which is fused with an LSTM-based estimator through an error-aware multi-estimator adaptive fusion strategy. Finally, PLA for drone RID frames is performed according to \textcolor{blue}{(\ref{Eq6})} and \textcolor{blue}{(\ref{Eq47})}. Since the deep learning-based detection and modulation parameter estimation of RID frames have been thoroughly investigated in our previous work\textcolor{blue}{\cite{ref_syn}}, it is assumed that the modulation parameter prior knowledge and a coarse time-frequency localization of the RID frame are already available.

\subsection{RID-Aware Wireless Sensing and Decoding Module}
\begin{enumerate} [leftmargin=0pt, itemindent=2pc, listparindent=\parindent]
	
	\item \textit{Estimation of \(\theta\) and \(\varphi\)}: The RID frames are sampled at the BS with a sampling rate of \(f_s\). A spectrum energy-based method is first employed to obtain a coarse estimate of \(f_c\), after which the received signal is shifted in the frequency domain to approximately zero center frequency. The signal is then passed through a low-pass filter and resampled with a rate of \(\frac{B}{f_s}\) to obtain the time-domain signal sequence \(x_{t,n}\), where \(n \in \mathcal{N} \triangleq \{1, 2, \ldots, \varpi f_s\}\).
	
	Then coarse time synchronization is performed to determine the start and end positions of the RID frames, which can be expressed as
	\begin{equation}
		\label{Eq7}
		M(n)
		= \frac{
			\left|
			\sum\limits_{m=1}^{K_1}
			x_{t,n+m}\, x^{*}_{t,n+m+K_2}
			\right|^{2}
		}{
			\left|
			\sum\limits_{m=1}^{K_1}
			x_{t,n+m}\, x^{*}_{t,n+m}
			\right|
			\left|
			\sum\limits_{m=1}^{K_1}
			x_{t,n+m+K_2}\, x^{*}_{t,n+m+K_2}
			\right|
		}. 
	\end{equation}
	
	When \(K_1=N_\text{cp}\) and \(K_2=N\), time coarse synchronization is achieved by exploiting the CP, and the index of the first prominent peak of \(M(n)\) is identified as the frame start position. When \(K_1=K_2=\frac{N}{2}\), time coarse synchronization is performed using the first ZC sequence in the frame, and the frame start position is obtained by subtracting \(2N+3N_\text{cp}\) from the index of the peak of \(M(n)\).

	For \(x_{t,n}\) with the start and end positions already estimated, the estimation of \(\theta\) and \(\phi\) can be expressed as
	\begin{equation}
		\label{Eq8}
		(\hat{\theta}, \hat{\varphi})
		= \underset{\theta, \varphi}{\arg \max}\,
		\left| \mathbf{a}^{H}(\theta, \varphi)\, \mathbf{X}_{i} \right|. 
	\end{equation}
	where the resolution of \(\theta \in [0, 2\pi]\) is \(\frac{1.772}{N_\text{rx}\cos\varphi}\) and that of \(\varphi \in \left[-\frac{\pi}{2}, \frac{\pi}{2}\right]\) is \(\frac{1.772}{N_\text{ry}}\), both measured in radians\textcolor{blue}{\cite{ref_aoa}}.

	\item \textit{Estimation of \(f_d(t)\)}: In wireless communication systems, a carrier frequency mismatch commonly exists between the transmitter and the receiver, mainly due to oscillator frequency discrepancies and Doppler shifts caused by mobility. The former yields a constant offset, while the latter produces a time-varying offset \(f_d(t)\) whose dynamics depend on the relative velocity \(v(t)\). For drone A2G channels, the overall frequency offset is typically confined within \(\frac{B}{N}\), making fine frequency synchronization is sufficient to estimate the normalized frequency offset \(\varepsilon(t)\) and subsequently obtain \(\hat{f}_d(t)\).
	
	Let \(R_\text{max}(t)\) denote the auto-correlation value at the peak of \(\left|\sum_{m=1}^{K_1}x_{t,n+m}\, x^{*}_{t,n+m+K_2}\right|\), then we have  
	\begin{equation}
		\label{Eq9}
		\hat{\varepsilon}(t) = - \arctan\!\left(\frac{\operatorname{Im}\!\left(R_{\max}(t)\right)} {\operatorname{Re}\!\left(R_{\max}(t)\right)}\right)\times
		\frac{N}{2\pi K_1}. 
	\end{equation}

	Therefore, the Doppler frequency shift can be estimated by
	\begin{equation}
		\label{Eq10}
		\hat{f}_d(t) \approx \frac{B}{N}\hat{\varepsilon}(t). 
	\end{equation}

	\item \textit{Estimation of ACG}: After completing fine frequency synchronization via \textcolor{blue}{(\ref{Eq9})}, fine time synchronization is performed to accurately determine the symbol boundaries within the RID frame and to remove all CP. To mitigate the impact of the wireless channel on \(x_{t,n}\), channel estimation and equalization are further carried out prior to decoding. Since the RID frame contains two distinct ZC sequence symbols, let \(\mathbf{s}^1_t\) and \(\mathbf{s}^2_t\) denote the first and second transmitted ZC sequence symbols, respectively, and let \(\mathbf{x}^1_t\) and \(\mathbf{x}^2_t\) denote the corresponding received ZC sequence symbols. The channel impulse response can then be estimated as
	\begin{equation}
		\label{Eq11}
		\hat{\mathbf{h}}_t
		= \underset{\mathbf{h}}{\arg \max}\,
		\sum_{i=1}^{2}
		\left|
		\bigl( \mathbf{x}^i_t - \mathbf{s}^i_t \mathbf{h}_t \bigr)^{\dagger}
		\bigl( \mathbf{x}^i_t - \mathbf{s}^i_t \mathbf{h}_t \bigr)
		\right|.
	\end{equation}
	
	Channel equalization is then performed using \(\hat{\mathbf{h}}_t^{-1}\), followed by demodulation and quantization to obtain the bitstream. Moreover, direct computation of the RSS incurs relatively high complexity, whereas \(\hat{\mathbf{h}}_t\) inherently captures the distance between the drone and the BS and is already available from the decoding process without introducing additional overhead. Therefore, the ACG can be computed as \(\mathbb{E}\!\left[\left\|\hat{\mathbf{h}}_{t}\right\|_{2}^{2}\right]\).

	\item \textit{Estimation of \(N_t\)}: Let \(\tilde{\mathbf{X}}_i(t)\) denote the frequency-domain representation of the RID frame after time-frequency synchronization, the covariance matrix on \(i\)-th subcarrier can be computed as
	\begin{equation}
		\label{Eq12}
		\mathbf{R}_i(t) = \frac{1}{N_{\text{sym}}} \tilde{\mathbf{X}}_i(t) \tilde{\mathbf{X}}_i^{\dagger}(t).
	\end{equation}
	
	Moreover, \(\mathbf{R}_i(t)\) can be decomposed into eigenvalues to obtain
	\begin{equation}
		\label{Eq13}
		l^t_{1,i} \geqslant l^t_{2,i} \geqslant \ldots l^t_{N_t,i} \geqslant l^t_{N_t+1,i} \geqslant \ldots \geqslant l^t_{N_r,i}.
	\end{equation}
	
	The objective function of estimating \(\hat{N}_t\) can be expressed as\textcolor{blue}{\cite{ref_nt}}
	\begin{equation}
		\label{Eq14}
		\hat{N}_t=\underset{j=0,1,\ldots,N_r-1}{\mathop{\arg \min }}\,\sum\limits_{i=1}^{N_\text{dv}}{\text{AIC}_i(j)},
	\end{equation}
	where \(\text{AIC}_i(j) = -2(N_r-k)N_\text{sym}\log\!\left(\frac{\mathbb{G}\!\left[\mathbf{l}^{t}_{j+1,i}\right]} {\mathbb{E}\!\left[\mathbf{l}^{t}_{j+1,i}\right]}\right)+2k(N_r-k)\) and \(\mathbf{l}^t_{j,i}=\bigl[ l^{t}_{j+1,i}, \ldots, l^{t}_{N_r,i} \bigr]\).
	 
	 \item \textit{RID Frame Decoding}: The bitstream is first descrambled using the predefined Gold sequences to eliminate the effect of transmitter scrambling, and is then processed by a Turbo decoder. The decoded bits are subsequently demapped to recover the payload information carried in the RID frame. The parameter vector extracted from the RID frame for PLA can be expressed as
	 \begin{equation}
	 	\label{Eq16}
	 	\mathbf{y} = \bigl[\mathbf{P}_u,v_f,v_r,v_d,a_f,a_r,a_d,\psi,\omega\bigr],
	 \end{equation}
	 where \(v_f\), \(v_r\), \(v_d\), and \(\psi\) denote the forward velocity, rightward velocity, downward velocity, and yaw angle of the drone in the body frame. The acceleration vector \(\bigl[a_f,a_r,a_d\bigr]\) and the yaw rate \(\omega\) can be obtained from the differences between adjacent time slots. In addition, the drone’s UUID and type can be obtained, thereby acquiring \(N_t\) corresponding to the drone type.
\end{enumerate}

\subsection{Communication-Aware Motion State Estimator}

The kinematic estimator is formulated as a yaw-augmented CA-EKF. Unlike conventional CA model defined purely in the world frame, the proposed estimator represents translational acceleration in the body frame and explicitly incorporates yaw and turn-rate dynamics into the state vector. This design enables accurate motion estimator for both straight-line flight and moderate maneuvering, while preserving the simplicity and numerical stability of the CA models. By coupling body-frame acceleration with heading evolution, the proposed yaw-augmented CA-EKF provides a unified and physically consistent kinematic representation of translational and rotational motion, thereby serving as a reliable motion prior for subsequent state estimation and fusion..

\begin{enumerate}[leftmargin=0pt, itemindent=2pc, listparindent=\parindent]
	\item{\textit{Motion Parameter Construction from Sensing Measurements}}: To map ACG to \(d(t)\), a lightweight regression network is designed for scalar ACG inputs. Specifically, a logarithmic transformation is first applied to the ACG to compress large-magnitude values while amplifying relative differences among small-magnitude values, thereby enhancing sensitivity in the low-amplitude regime without significantly distorting large-scale variations. The resulting log-transformed feature is then fed into a single fully connected layer with the learnable weight \(\mathbf{\Theta}\) and bias \(b\), which performs an affine mapping to directly estimate \(d(t)\). Owing to its log-linear formulation, the proposed network contains only two trainable parameters, enabling stable training and strong generalization performance despite the extremely low-dimensional input. The mapping process can be expressed as
	\begin{equation}
		\label{Eq17}
		\hat{d}(t) = \mathbf{\Theta}\log\!\left(\mathbb{E}\!\left[\left\|\hat{\mathbf{h}}_{t}\right\|_{2}^{2}\right]\right) + b,
	\end{equation}
	
	Then the sensing drone position can be computed by
	\begin{equation}
		\label{Eq18}
		\tilde{\mathbf{P}}_u(t) = \mathbf{P}_b + \hat{d}(t) \begin{bmatrix}
			\cos \hat{\varphi} \cos \hat{\theta} \\
			\cos \hat{\varphi} \sin \hat{\theta} \\
			\sin \hat{\varphi}
		\end{bmatrix},
	\end{equation}
	and the sensing drone velocity can be computed by
	\begin{equation}
		\label{eqv}
		\tilde{v}(t) = \frac{c\hat{f}_d(t)}{f_c(t)\cos \theta(t)}.
	\end{equation}
	
	\item{\textit{Yaw-Augmented CA-EKF}}: To ensure consistency between the motion parameters extracted in the body frame from the RID frames and the world-frame state representation adopted in the kinematic model, the velocity and acceleration are mapped from the body frame to the world frame. Although the velocity states are defined in the body frame, the signal-derived speed measurement resides in the world frame. By explicitly embedding the body-to-world transformation into the nonlinear measurement function, the proposed EKF enforces cross-frame consistency without requiring state reparameterization, which can be expressed as follows.
	\begin{equation}
		\label{Eq19}
		\mathbf{v}
		=
		\begin{bmatrix}
			v_x \\ v_y \\ v_z
		\end{bmatrix}
		=
		\begin{bmatrix}
			\mathbf{R}(\psi)
			\begin{bmatrix}
				v_f \\ v_r
			\end{bmatrix} \\
			- v_d
		\end{bmatrix}
		=
		\begin{bmatrix}
			\begin{bmatrix}
				\cos \psi & -\sin \psi \\
				\sin \psi & \cos \psi
			\end{bmatrix}
			\begin{bmatrix}
				v_f \\ v_r
			\end{bmatrix} \\
			- v_d
		\end{bmatrix}.
	\end{equation}
	\begin{equation}
		\label{Eq20}
		\mathbf{a}
		=
		\begin{bmatrix}
			a_x \\ a_y \\ a_z
		\end{bmatrix}
		=
		\begin{bmatrix}
			\mathbf{R}(\psi)
			\begin{bmatrix}
				a_f \\ a_r
			\end{bmatrix} \\
			- a_d
		\end{bmatrix}.
	\end{equation}
	
	The updates of the drone position and velocity can be expressed as follows, respectively.
	\begin{equation}
		\label{Eq21}
		\mathbf{p}_{t+1}=\mathbf{p}_t+{\mathbf{v}}_t\varpi +\frac{1}{2}\mathbf{a}_t{\varpi }^2.
	\end{equation}
	\begin{equation}
		\label{Eq22}
		\mathbf{v}_{t+1}=\mathbf{v}_t+\mathbf{a}_t\varpi,
	\end{equation}
	where \(\mathbf{a}_{t+1}=\mathbf{a}_t\).
	
	The update of the drone yaw angle can be expressed as
	\begin{equation}
		\label{Eq23}
		\psi_{t+1}=\psi _t+\omega_t\varpi,
	\end{equation}
	where \(\omega_{t+1}=\omega_t\).
	
	The EKF is adopted since \textcolor{blue}{(\ref{Eq21})} is nonlinear, and the corresponding state transition Jacobian matrix \(\mathbf{F}\) can be obtained by
	\begin{equation}
		\label{Eq24}
		\mathbf{F}
		=
		\begin{bmatrix}
			\mathbf{F}_{pp} & \mathbf{F}_{pv} & \mathbf{F}_{pa} & \mathbf{F}_{p\psi} & \mathbf{F}_{p\omega} \\
			\mathbf{0}_{3\times 3} & \mathbf{F}_{vv} & \mathbf{F}_{va} & \mathbf{F}_{v\psi} & \mathbf{F}_{v\omega} \\
			\mathbf{0}_{3\times 3} & \mathbf{0}_{3\times 3} & \mathbf{F}_{aa} & \mathbf{F}_{a\psi} & \mathbf{F}_{a\omega} \\
			\mathbf{0}_{1\times 3} & \mathbf{0}_{1\times 3} & \mathbf{0}_{1\times 3} & F_{\psi\psi} & F_{\psi\omega} \\
			\mathbf{0}_{1\times 3} & \mathbf{0}_{1\times 3} & \mathbf{0}_{1\times 3} & 0 & F_{\omega\omega}
		\end{bmatrix}
		\in \mathbb{R}^{11 \times 11},
	\end{equation}
	where \(\mathbf{F}_{pp} = \mathbf{F}_{vv} = \mathbf{F}_{va} = \mathbf{F}_{aa} = \mathbf{I}_3\), \(\mathbf{F}_{p\omega} = \mathbf{F}_{v\psi} = \mathbf{F}_{v\omega}
	= \mathbf{F}_{a\psi} = \mathbf{F}_{a\omega} = \mathbf{0}_{3\times 1}\), \(F_{\psi\psi} = F_{\omega\omega} = 1\), \(F_{\psi\omega} = \varpi\),
	\begin{equation}
		\label{Eq25}
		\mathbf{F}_{pa}
		= \frac{\varpi}{2} \mathbf{F}_{pv}
		= \frac{\varpi^{2}}{2}
		\begin{bmatrix}
			\cos \psi & -\sin \psi & 0 \\
			\sin \psi & \cos \psi  & 0 \\
			0         & 0          & -1
		\end{bmatrix},
	\end{equation}
	and
	\begin{equation}
		\label{Eq26}
		\mathbf{F}_{p\psi}\text{=}\begin{bmatrix}
			\bigl(-v\!_f\!\sin\psi\!-\!v_r\!\cos\psi\bigr)\varpi\!-\!\bigl(a\!_f\!\sin\psi\!+\!a_r\!\cos\psi\bigr)\frac{\varpi^2}{2}  \\
			\bigl(v\!_f\!\cos\psi\!-\!v_r\!\sin\psi\bigr)\varpi\!+\!\bigl(a\!_f\!\cos\psi\!-\!a_r\!\sin\psi\bigr)\frac{\varpi^2}{2}  \\
			0
			\end{bmatrix}.
	\end{equation}
	
	When both high-confidence historical decoding information and real-time sensing measurements are available at the same time slot, a sequential EKF update strategy is adopted. Specifically, the previously authenticated motion state is first exploited to strongly constrain the predicted state, after which real-time sensing-derived motion information is incorporated in a second update stage to refine the prediction.
	
	Based on the yaw-augmented CA motion model, the current state vector is estimated from the previously validated state and can be expressed as
	\begin{equation}
		\label{Eq27}
		\hat{\mathbf{y}}_{t|t-1} = \mathbf{F}\hat{\mathbf{y}}_{t-1|t-1},
	\end{equation}
	
	The uncertainty of the state estimation can be characterized by the estimation error covariance matrix, which can be computed by
	\begin{equation}
		\label{Eq28}
		\mathbf{E}_{t|t-1} = \mathbf{F} \mathbf{E}_{t-1|t-1} \mathbf{F}^{\mathsf{T}} + \mathbf{Q}_t,
	\end{equation}
	where \(\mathbf{Q}_t\) denotes the process noise covariance matrix. Since the pseudo-measurement state vector \(\hat{\mathbf{z}}^{1}_{t|t-1}\) is constructed as a linear subset of \(\hat{\mathbf{y}}_{t|t-1}\) with added observation noise, the measurement Jacobian matrix can be expressed as
	\begin{equation}
		\label{Eq29}
		\mathbf{M}
		=
		\begin{bmatrix}
			\mathbf{I}_3 & \mathbf{0}_{3\times 3} & \mathbf{0}_{3\times 3} & \mathbf{0}_{3\times 1} & \mathbf{0}_{3\times 1} \\
			\mathbf{0}_{3\times 3} & \mathbf{I}_3 & \mathbf{0}_{3\times 3} & \mathbf{0}_{3\times 1} & \mathbf{0}_{3\times 1} \\
			\mathbf{0}_{1\times 3} & \mathbf{0}_{1\times 3} & \mathbf{0}_{1\times 3} & 1 & 0
		\end{bmatrix}.
	\end{equation}
	
	Then the observation residual can be computed by
	\begin{equation}
		\label{Eq30}
		\mathbf{r}^{1}_{t} = \hat{\mathbf{z}}^{1}_{t|t-1} - \mathbf{M} \hat{\mathbf{y}}_{t|t-1},
	\end{equation}
	
	Since the filter operates in the steady state where \(\mathbb{E}\!\left[ \mathbf{r}_t \mathbf{r}_t^{\mathsf{T}} \right] = \mathbf{M} \mathbf{P}_{t|t-1} \mathbf{M}^{\mathsf{T}} + \mathbf{R}_t\) holds, the observation noise is updated via covariance matching estimation. The observation residual covariance matrix can be updated by
	\begin{equation}
		\label{Eq31}
		\mathbf{S}_t = \alpha \mathbf{S}_{t-1} + (1 - \alpha) \mathbf{r}_t \mathbf{r}_t^{\mathsf{T}},
	\end{equation}
	where \(\alpha\) is a weighting factor. 
	
	The observation noise covariance matrix can be updated by
	\begin{equation}
		\label{Eq32}
		\mathbf{R}_t = \text{PSD}\Bigl(\mathbf{S}_t - \mathbf{M} \mathbf{E}_{t|t-1} \mathbf{M}^{\mathsf{T}}\Bigr),
	\end{equation}
	where \(\text{PSD}\) denotes the positive semi-definite projection operator that enforces the symmetry and positive semi-definiteness of the matrix. The Kalman gain can be computed as
	\begin{equation}
		\label{Eq33}
		\mathbf{K}_t = \mathbf{E}_{t|t-1} \mathbf{M}^{\mathsf{T}} \Bigl( \mathbf{M} \mathbf{E}_{t|t-1} \mathbf{M}^{\mathsf{T}} + \mathbf{R}_t \Bigr)^{-1},
	\end{equation}
	
	The estimated state can be updated by
	\begin{equation}
		\label{Eq34}
		\hat{\mathbf{y}}^{1}_{t|t} = \hat{\mathbf{y}}_{t|t-1} + \mathbf{K}_t \mathbf{r}_t,
	\end{equation}
	
	The process residual can be computed as
	\begin{equation}
		\label{Eq35}
		\mathbf{w}^{1}_{t} = \hat{\mathbf{y}}^{1}_{t|t} - \hat{\mathbf{y}}_{t|t-1},
	\end{equation}
	
	The process noise can be updated by
	\begin{equation}
		\label{Eq36}
		\mathbf{Q}_t = \text{PSD}\bigl( \mathbf{W}_t \bigr) = \text{PSD}\Bigl( \beta \mathbf{W}_{t-1} + (1-\beta) \mathbf{w}_t \mathbf{w}_t^{\mathsf{T}} \Bigr),
	\end{equation}
	where \(\beta\) is a weighting factor. 
	
	The covariance matrix can be updated as
	\begin{equation}
		\label{Eq37}
		\begin{aligned}
			\mathbf{E}_{t|t}
			&= (\mathbf{I} - \mathbf{K}_t \mathbf{M}) \mathbf{E}_{t|t-1}  (\mathbf{I} - \mathbf{K}_t \mathbf{M})^{\mathsf{T}} + \mathbf{K}_t \mathbf{R} \mathbf{K}_t^{\mathsf{T}} \\
			&= (\mathbf{I} - \mathbf{K}_t \mathbf{M}) \mathbf{E}_{t|t-1}.
		\end{aligned}
	\end{equation}
	
	After completing an initial update based on historical high-dimensional pseudo-observations that provide strong constraints, a second update is performed by exploiting the overall velocity estimated from the current time-slot signal as a weak constraint. The state vector \(\hat{\mathbf{z}}^{2}_{t|t-1}\) estimated from the current time-slot signal can be expressed as
	\begin{equation}
		\label{Eq38}
		\hat{\mathbf{z}}^{2}_{t|t-1} = \bigl[ \tilde{\mathbf{P}}_u(t), \tilde{v}(t) \bigr].
	\end{equation}
	
	Accordingly, the observation residual \(\mathbf{r}^2_t\) is computed based on \(\hat{\mathbf{z}}^{2}_{t|t-1}\), \textcolor{blue}{(\ref{Eq19})}, and \textcolor{blue}{(\ref{Eq30})}. The corresponding measurement Jacobian matrix is given by
	\begin{equation}
		\label{Eq39}
		\mathbf{M}
		=
		\begin{bmatrix}
			\mathbf{I}_3 & \mathbf{0}_{3\times 3} & \mathbf{0}_{3\times 3} & \mathbf{0}_{3\times 1} & \mathbf{0}_{3\times 1} \\
			\mathbf{0}_{1\times 3} & \mathbf{M}_v & \mathbf{0}_{1\times 3} & M_{v\psi} & 0
		\end{bmatrix}\in \mathbb{R}^{4 \times 11},
	\end{equation}
	where \( \mathbf{M}_v = \frac{\begin{bmatrix} v_x\cos\psi+v_y\sin\psi & -v_x\sin\psi+v_y\cos\psi & v_d \end{bmatrix}}{v} \) and \(M_{v\psi} = \frac{v_x(-v_f\sin\psi-v_r\cos\psi)+v_y(v_f\cos\psi-v_r\sin\psi)}{v}\).
	
	The state estimation is carried out according to \textcolor{blue}{(\ref{Eq31})}--\textcolor{blue}{(\ref{Eq37})} to obtain the estimated position \(\hat{\mathbf{P}}^\text{EKF}_u(t)\) produced by the yaw-augmented CA-EKF.

\end{enumerate}

\subsection{Error-Aware Multi-Estimator Adaptive Fusion Strategy}
The yaw-augmented CA-EKF is built upon a constant-acceleration assumption and is therefore limited in modeling highly nonlinear motion behaviors, particularly under complex drone maneuvers and aggressive flight dynamics. In addition, its estimation performance mainly depends on short-term state propagation and instantaneous observations, which restricts its capability to exploit long-term trajectory dependencies.

To overcome these limitations, a data-driven LSTM-based motion estimator is introduced and adaptively fused with the model-driven yaw-augmented CA-EKF. By leveraging historical sequences of position, velocity, and yaw states, the LSTM network learns implicit long-term dynamic patterns that are difficult to capture with analytical kinematic models. Furthermore, a multi-head self-attention mechanism is incorporated to selectively emphasize informative temporal features. Based on the estimated error statistics of both estimators, an error-aware adaptive fusion strategy is employed to refine the CA-EKF predictions online, thereby significantly improving tracking robustness and accuracy under highly maneuvering and non-stationary drone motion conditions.

\begin{enumerate}[leftmargin=0pt, itemindent=2pc, listparindent=\parindent]
	\item{\textit{LSTM Network Architecture}}:
	The drone state vectors from the RID frames over the past \(T_1\) time slots are reshaped into an input vector \(\mathbf{U}_t = \bigl[\mathbf{y}_{t-T_1};\mathbf{y}_{t-T_1+1};\ldots;\mathbf{y}_{t-1}\bigr]\in \mathbb{R}^{T_1 \times 11}\). \(\mathbf{U}_t\) is then fed into an LSTM layer to capture short-term temporal dependency information, the element in the output vector \(\bm{\xi}_t \in \mathbb{R}^{T_1\times 128}\) of LSTM layer can be expressed as
	\begin{equation}
		\label{Eq41}
		\xi_j = \mathbf{o}_j \odot \tanh \Bigl( \mathbf{f}_j \odot \mathbf{c}_{j-1} + \mathbf{i}_j \odot \tilde{\mathbf{c}}_j \Bigr), \quad j \in \{t-T_1, \ldots, t-1\},
	\end{equation}
	where \(\mathbf{o}_j\), \(\mathbf{f}_j\), and \(\mathbf{i}_j\) denote the outputs of the output gate, forget gate, and input gate, respectively, while \(\mathbf{c}_{j-1}\) and \(\tilde{\mathbf{c}}_j\) represent the memory cell and candidate memory cell, respectively.
	
	The multi-head self-attention module is designed to enable the network to capture diverse temporal dependency patterns, thereby identifying the historical time steps most critical for predicting future drone trajectories and enhancing the modeling of long-range dependencies. A total of \(M_1=4\) attention heads are employed to compute the similarity between time steps within the sequence, and the output of \(m\)-th head can be expressed as
	\begin{equation}
		\label{Eq42}
		\mathbf{O}_m = \text{Softmax}\Biggl(
		\frac{ (\bm{\xi}\, \mathbf{\Theta}^Q_m)(\bm{\xi} \mathbf{\Theta}^K_m)^{\mathsf{T}} }{\sqrt{32}}
		\Biggr)\, (\bm{\xi}\mathbf{\Theta}^V_m)
		\in \mathbb{R}^{T_1 \times T_1},
	\end{equation}
	where \(\mathbf{\Theta}^Q_m\), \(\mathbf{\Theta}^K_m\), and \(\mathbf{\Theta}^V_m\) denote learnable projection matrices \(\in \mathbb{R}^{128 \times \frac{128}{M_1}}\) of the neural network. Then \(\mathbf{O}_m\) are concatenated along the second feature dimension and projected to obtain the output feature \(\mathbf{O}\in \mathbb{R}^{T_1 \times 128}\).
	
	\(\bm{\xi}_t\) and \(\mathbf{O}\) are combined through a residual connection to obtain \(\tilde{\mathbf{O}}_t\), allowing the attention mechanism to learn incremental information while preserving the original LSTM representations. \(\tilde{\mathbf{O}}\) is then processed sequentially by a normalization layer to improve network stability, a temporal average pooling layer with ReLU activation for feature compression, and a fully connected layer for predicting the future 3D position \(\hat{\mathbf{P}}^\text{LSTM}_u(t)\), which can be expressed as
	\begin{equation}
		\label{Eqlstm}
		\hat{\mathbf{P}}^{\text{LSTM}}_u(t)
		=
		\mathrm{FC}
		\!\left(
		\mathrm{ReLU}
		\!\left(
		\mathrm{AvgPool}
		\!\left(
		\mathrm{Norm}
		\!\left(
		\tilde{\mathbf{O}}_t
		\right)
		\right)
		\right)
		\right).
	\end{equation}
	
	The loss function is given by
	\begin{equation}
		\label{Eq43}
		\mathcal{L} = \frac{1}{T_\text{Tr}} \sum_{t=1}^{T_\text{Tr}} \left\| \hat{\mathbf{P}}^\text{LSTM}_u(t) - \mathbf{P}_u(t) \right\|_2^2,
	\end{equation}
	where \(T_\text{Tr}\) denotes the total number of time slots across the entire training process. 
	\item{\textit{Adaptive Fusion}}: The estimation error covariance matrix \(\mathbf{E}^{\text{LSTM}}_t\) for the LSTM-based motion estimator can be expressed as
	\begin{equation}
		\label{Eq44}
		\mathbf{E}^{\text{LSTM}}_t
		= \frac{1}{T_{\text{Tr}}} \sum_{i=1}^{T_{\text{Tr}}} 
		\bigl( \hat{\mathbf{P}}^\text{LSTM}_u(i) - \mathbf{P}_u(i) \bigr)
		\bigl( \hat{\mathbf{P}}^\text{LSTM}_u(i) - \mathbf{P}_u(i) \bigr)^{\mathsf{T}}.
	\end{equation}
	
	\(\mathbf{E}^{\text{LSTM}}_0\) is precomputed offline using the training dataset and continuously updated during online inference. The estimation error covariance matrix \(\mathbf{E}^{\mathrm{EKF}}_t\) for the yaw-augmented CA-EKF-based motion estimator can be obtained from \(\mathbf{E}_{t|t-1}\). And the error-aware multi-estimator adaptive fusion can be expressed as
	\begin{equation}
		\label{Eq45}
		\hat{\mathbf{P}}^{\mathrm{Fused}}_u(t)
		= \hat{\mathbf{P}}^{\mathrm{EKF}}_u(t)
		+ \alpha_t
		\bigl( \hat{\mathbf{P}}^{\mathrm{LSTM}}_u(t) - \hat{\mathbf{P}}^{\mathrm{EKF}}_u(t) \bigr).
	\end{equation}
	where \(\alpha_t = \mathbf{E}^{\mathrm{EKF}}_t \bigl( \mathbf{E}^{\mathrm{EKF}}_t + \mathbf{E}^{\mathrm{LSTM}}_t \bigr)^{-1}\).
\end{enumerate}

	\subsection{NFZ Constraints}
	Since NFZs formed by airports, military areas, and dense urban structures impose strict constraints on legal drone trajectories, determining whether a drone’s continuous trajectory intersects an NFZ is of critical importance for drone RID frame PLA. The continuous horizontal trajectory \(\mathbf{e}^\delta_u(t)\in \mathbb{R}^2\) of a drone between two consecutive time slots can be expressed as
	\begin{equation}
		\label{Eq46}
		\mathbf{e}^\delta_u(t) = (1-\delta)\mathbf{e}_u(t)+\delta\mathbf{e}_u(t+1),\quad \delta \in [0,1].
	\end{equation}
	
	Let \(\mathcal{P} \triangleq \{\bm{\tau}_1, \bm{\tau}_2, \ldots, \bm{\tau}_{M_2}\}\subset \mathbb{R}^{M_2\times2}\) denote the set of vertices defining a simple polygonal, whose boundary is closed and non-self-intersecting, with the vertices arranged in a consistent clockwise order, and \(\bm{\tau}_{M_2+1} \triangleq \bm{\tau}_1\). If the drone trajectory intersects the NFZ, the following global topological criterion based on the winding number holds:
	\begin{equation}
		\label{Eq47}
		\left|
		\frac{1}{2\pi}
		\sum_{i=1}^{M_2}
		\operatorname{atan2}
		\!\left(
		\bm{\varsigma}_{1,i} \times \bm{\varsigma}_{2,i},\,
		\bm{\varsigma}_{1,i} \cdot \bm{\varsigma}_{2,i}
		\right)
		\right|
		\ge 1,
	\end{equation}
	where \(\bm{\varsigma}_{1,i} = \bm{\tau}_i - \mathbf{e}^\delta_u(t)\), \(\bm{\varsigma}_{2,i} = \bm{\tau}_{i+1} - \mathbf{e}^\delta_u(t)\), and \(\text{atan2}(\cdot,\cdot)\) denotes the four-quadrant inverse tangent function.
	
	It is worth noting that if \(\mathbf{e}^\delta_u(t)\) coincides exactly with a vertex \(\bm{\tau}_i\), the two vectors \(\bm{\varsigma}_{1,i}\) and \(\bm{\varsigma}_{2,i}\) become collinear with zero magnitude, yielding
	\begin{equation}
		\bm{\varsigma}_{1,i} \times \bm{\varsigma}_{2,i} = \bm{\varsigma}_{1,i} \cdot \bm{\varsigma}_{2,i} = 0,
	\end{equation}
	which renders the \(\text{atan2}(\cdot,\cdot)\) operation undefined. Moreover, when \(\mathbf{e}^\delta_u(t)\) lies on the boundary of the NFZ, floating-point round-off errors may lead to numerical instability in the accumulated winding angle. Therefore, to ensure robust and reliable NFZ constraint determination, it is necessary to explicitly check whether \(\mathbf{e}^\delta_u(t)\) coincides with any vertex or lies on any edge of the NFZ prior to applying \textcolor{blue}{(\ref{Eq47})}.
	
	The core details of the proposed PLA algorithm for drone RID frames are summarized in Algorithm \ref{algPLA}. It is worth noting that if $\hat{N}_t$ does not match the expected $N_t$ associated with the decoded drone type, the corresponding drone is regarded as illegal. Likewise, if the predicted trajectory violates the NFZ constraints, the drone is also deemed illegal. The PLA outcome is represented by a binary indicator \(D_t\), where \(D_t=1\) indicates that the authentication is successfully passed and the drone is deemed legitimate, and \(D_t=0\) otherwise.
	
\begin{algorithm}[H]
	\caption{PLA Algorithm for Drone RID Frames}
	\label{algPLA}
	\begin{algorithmic}[1]
		
		\State \textbf{Input:} $f_s$ and $\mathbf{S}_i(t)$.
		
		\State \textbf{Output:} \(D_t\).
		
		\State Initialize $\mathbf{E}_0$, $\mathbf{Q}_0$, $\mathbf{S}_0$, $\mathbf{R}_0$, and $\mathbf{W}_0$.
		
		\State \textbf{for} \(t = 1:T\) \textbf{do}
		
		\State \hspace{0.5cm} Estimate $\hat{\theta}$, $\hat{\varphi}$, $\hat{f}_d(t)$, $\hat{\mathbf{h}}_t$, and $\hat{N}_t$ by \textcolor{blue}{(\ref{Eq8})}, \textcolor{blue}{(\ref{Eq10})}, \textcolor{blue}{(\ref{Eq11})},
		\Statex \hspace{0.5cm} and \textcolor{blue}{(\ref{Eq14})}.
		
		\State \hspace{0.5cm} Obtain the decoding information \(\mathbf{y}\) according to \textcolor{blue}{(\ref{Eq16})}.
		
		\State \hspace{0.5cm} Construct the decoding parameter vector \(\hat{\mathbf{y}}_{t|t-1}\) and
		\Statex \hspace{0.5cm} $\hat{\mathbf{z}}^{1}_{t|t-1}$ by \textcolor{blue}{(\ref{Eq27})}.
		
		\State \hspace{0.5cm} Update the estimation error covariance $\mathbf{E}_{t|t-1}$ by \textcolor{blue}{(\ref{Eq28})}.
		
		\State \hspace{0.5cm} Update the residual covariance $\mathbf{S}_t$ according to \textcolor{blue}{(\ref{Eq31})}.
		
		\State \hspace{0.5cm} Update the observation noise covariance $\mathbf{R}_t$ by \textcolor{blue}{(\ref{Eq32})}.
		
		\State \hspace{0.5cm} Perform the initial EKF update to obtain $\hat{\mathbf{y}}^{1}_{t|t}$ by \textcolor{blue}{(\ref{Eq34})}.
		
		\State \hspace{0.5cm} Update the process noise $\mathbf{Q}_t$ based on \textcolor{blue}{(\ref{Eq36})}.
		
		\State \hspace{0.5cm} Update  $\mathbf{E}_{t|t}$ using \textcolor{blue}{(\ref{Eq37})}.
		
		\State \hspace{0.5cm} Construct the sensing parameter vector $\hat{\mathbf{z}}^{2}_{t|t-1}$ by \textcolor{blue}{(\ref{Eq38})}.
		
		\State \hspace{0.5cm} Perform the second EKF update by reusing Steps~9--
		\Statex \hspace{0.5cm} 13, and obtain the EKF-based position estimate
		\Statex \hspace{0.5cm} $\hat{\mathbf{P}}^{\mathrm{EKF}}_u(t)$.
		
		\State \hspace{0.5cm} Obtain the LSTM-based position estimate $\hat{\mathbf{P}}^{\mathrm{LSTM}}_u(t)$
		\Statex \hspace{0.5cm} via \textcolor{blue}{(\ref{Eqlstm})}.
		
		\State \hspace{0.5cm} Fuse multi-estimator to obtain $\hat{\mathbf{P}}^{\mathrm{Fused}}_u(t)$ based on \textcolor{blue}{(\ref{Eq45})}.
		
		\State \hspace{0.5cm} Validate $\hat{N}_t$ against $N_t$ determined by the drone type.
		
		\State \hspace{0.5cm} Check compliance with the NFZ constraints by \textcolor{blue}{(\ref{Eq47})}.
		
		\State \hspace{0.5cm} Conduct location consistency check via \textcolor{blue}{(\ref{Eq6})}.
		
		\State \textbf{end for}
		
	\end{algorithmic}
\end{algorithm}

\subsection{Convergence and Complexity Analysis}

Owing to the inherent stability properties of EKF, the filter converges to a consistent state estimate regardless of the initial state magnitude, provided that the system is locally observable and the noise statistics are properly bounded\textcolor{blue}{\cite{ref_ekf}}. In contrast, the stability of LSTM-based motion estimator is ensured in a data-driven manner through offline training on representative trajectories and online error-aware fusion with the EKF output. By being constrained by the physically consistent EKF prior and adaptively weighted according to its estimated error, the LSTM avoids divergence while effectively capturing long-term nonlinear motion patterns.

The computational complexity of sensing and decoding information extraction, communication-aware motion state estimation, and error-aware multi-model adaptive fusion is \(\mathcal{O}\!\left(
N\!\left(N_{\text{sym}}+\log N+N_r N_t+N_r N_t^{2}\right)+N_r^{3} \right)\), \(\mathcal{O}\!\left(4\cdot\left(11^{3}+4^{3}\right)\right)\), and \(\mathcal{O}\!\left(T_1\!\left(11\cdot128+128^{2}\right) +T_1^{2}\cdot128 +T_1\cdot128^{2} +12+M_2\right)\), respectively. Accordingly, the overall computational complexity of the proposed algorithm can be approximated as \(\mathcal{O}\!\Bigl( N\!\left(N_{\text{sym}}+\log N+N_r N_t+N_r N_t^{2}\right) + N_r^{3} + 128\,T_1^{2} + 34176\,T_1\Bigr).\)

\section{Numerical Results}
\label{sec4}

In this section, the numerical simulation results are shown to validate the feasibility and performance of the proposed drone RID frame PLA algorithm in A2G networks.

\subsection{Simulation Setup}

A 3D Cartesian coordinate system is considered, where the BS is located at (500,500,10) m. In scenario 1, a legitimate drone takes off from the origin (0,0,0) and follows a smooth curved trajectory toward the destination (3500,3500,120) over \(T=300\) time slots, which continuously transmits RID frames during each \(\varpi\). The drone acceleration is randomly generated at each time slot to emulate realistic non-uniform motion dynamics. In scenario 2, the legitimate drone first travels along a straight-line trajectory and then transitions into a spiral ascending motion. Similar to Scenario 1, the acceleration is randomly generated, and the drone periodically transmits RID frames at every time slot. In both scenarios 1 and 2, illegal drones randomly appear within the spatial region spanned by the legitimate drone’s departure and destination points. To conceal their malicious behaviors, the illegal drones transmit replayed RID frames intercepted from legitimate drones, thereby impersonating authorized devices at the signaling level. In Scenario 1, four square NFZs are deployed around the BS, each with a side length of 20 m. The trajectory of the legitimate drone is pre-planned to avoid all NFZs, whereas the randomly generated illegal drones in Scenario 1 may enter or traverse these restricted regions. 

Scenario 3 is based on a real-world RF signal dataset collected in an urban environment. Multiple types of drones are deployed, each taking off and flying over a certain distance. The received RF signals consist not only of RID frames, but also include uplink control signals, downlink communication signals, as well as coexisting wireless transmissions such as Wi-Fi and Bluetooth. This scenario is used to evaluate the robustness of the proposed algorithm under practical and heterogeneous RF interference conditions.

The hardware environment consists of an NVIDIA GeForce RTX 4060 Ti GPU and an 14th Gen Intel(R) Core(TM) i7-14700K @ 3.40GHz, with the remaining parameters specification are provided in Table \ref{tab:table_para}\textcolor{blue}{\cite{ref_343}},\textcolor{blue}{\cite{ref_334}}.

\begin{table}[!t]
	\caption{Simulation Parameters \label{tab:table_para}}
	\centering
	\begin{tabular}{c c c}
		\hline
		Parameter & Value & Explanation\\
		
		\hline
		\(T\) & 8, 15, 300 & Number of time slots\\
		\(\varpi\)  & 1 & The duration of each time slot\\
		
		\(N_t\)  & 1,2 & Number of transmit antennas\\
		\(N_r\) & \(8\times8\) & Number of transmit antennas\\
		\(N_\text{sym}\) & 8, 9 & OFDM symbols in the RID frames\\
		\(N\)  & 1024 & Number of subcarriers\\
		\(N_\text{dc}\)  & 600 & Number of data subcarriers\\
		\(N_\text{vc}\)  & 424 & Number of virtual subcarriers\\
		\(B\) & 15.36 MHz & Signal bandwidth\\
		\(f_c\) & 2.4 GHz, 5.8 GHz & Center transmission frequency\\
		\(N_\text{cp}\) & 72, 80 & Cyclic prefix\\
		\(\sigma_n^2\) & -100 dBm & AWGN noise power\\
		\(c\) & \(3\times 10^8\) m/s & Speed of light\\
		\(P_t\) & \(30\) dBm & Maximum transmit power\\
		\(G_t\) & \(2\) dB & Transmit antenna gain\\
		\(G_r\) & \(30\) dB & Receive antenna gain\\
		\(f_s\) & 100 MHz & Sampling rate\\
		\(\alpha\) & 0.2 & The weighting factor in \textcolor{blue}{(\ref{Eq31})}\\
		\(\beta\) & 0.3 & The weighting factor in \textcolor{blue}{(\ref{Eq36})}\\
		\(M_1\) & 4 & Number of attention heads\\
		\(\delta\) & [0,1] & The weighting factor in \textcolor{blue}{(\ref{Eq46})}\\
		\(M_2\) & 4 & Number of vertices of one NFZ\\
		\hline
	\end{tabular}
\end{table}

The baseline schemes are given as follows.
\begin{enumerate}[leftmargin=0pt, itemindent=2pc, listparindent=\parindent]
	\item{\textit{RSS-Based Signal Authentication (RSS)}}:
	The RSS of the received RID frames is used to infer the propagation distance, and a consistency check is performed by comparing the estimated distance with the location advertised in the RID frames to detect potential drone location spoofing attacks\textcolor{blue}{\cite{ref_268}}.
	
	\item{\textit{CFO-Based Signal Authentication (CFO)}}:
	The CFO is extracted using ZC sequence symbols, enabling the estimation of the drone velocity and the evaluation of the consistency between the channel-dependent motion characteristics at different locations and the velocity information broadcast in the RID frames\textcolor{blue}{\cite{ref_269}},\textcolor{blue}{\cite{ref_176}}.
	
	\item{\textit{SNR Difference-Based Signal Authentication (SNRD)}}:
	The SNR difference between adjacent time slots is computed, to identify potential impersonation or spoofing attacks caused by RID frames transmitted from different locations\textcolor{blue}{\cite{ref_372}}.
	
	\item{\textit{AoA and Channel Gain-Based Signal Authentication (AoA+CG)}}:
	The drone location is estimated using the channel gain and AoA, which is compared with the location broadcast in the RID frames to detect potential location spoofing attacks\textcolor{blue}{\cite{ref_270}},\textcolor{blue}{\cite{ref_339}}.
	
	\item{\textit{Angle Delay Power Spectrum-Based Signal Authentication (ADPS)}}:
	The angle delay power spectrum is derived from the received RID frames, and the temporal consistency of the channel-dependent propagation characteristics across adjacent time slots is examined to authenticate the drone against the information broadcast in the RID frames\textcolor{blue}{\cite{ref_276}}.
	
	\item{\textit{LSTM-Based Signal Authentication (LSTM)}}:
	The decoded information from the historical RID frames that have been authenticated based on RSS is fed into an LSTM network to estimate the drone location, which is then verified against the location information broadcast in the current time slot\textcolor{blue}{\cite{ref_307}}.
	
	\item{\textit{EKF-Based Signal Authentication (EKF)}}:
	Based on the historical decoded information and the real-time sensed measurements, an EKF is utilized for location estimation, which is then authenticated against the location information broadcast in the current RID frame\textcolor{blue}{\cite{ref_334}}.
	
	\item{\textit{Yaw-Augmented EKF-Based Signal Authentication (Yaw EKF)}}:
	The proposed yaw-augmented EKF estimator is used to jointly estimate and authenticate the drone location by exploiting the historical decoded information and the real-time sensed measurements, with respect to the location information broadcast in the current RID frame.
\end{enumerate}

The performance of the drone state estimation algorithms is evaluated using the root mean square error (RMSE) and the mean absolute percentage error (MAPE). These metrics are employed in a unified manner to quantify the estimation accuracy of different state variables, including position, velocity, and angle, by measuring the absolute and relative deviations between the estimated states and their ground-truth values. The RMSE and MAPE of location are defined as
\begin{equation}
	\label{Eq_metric1}
	\text{RMSE}=\sqrt{\mathbb{E}\!\left[\left\|\mathbf{P}_u(t) - \hat{\mathbf{P}}_u(t)\right\|_2^2\right]},
\end{equation}

\begin{equation}
	\label{Eq_metric2}
	\text{MAPE}=\mathbb{E}\!\left[
	\frac{
			\left\|
			\mathbf{P}_u(t) - \hat{\mathbf{P}}_u(t)
			\right\|_2
		}{
			\left\|
			\mathbf{P}_u(t)
			\right\|_2
		}
		\right]
		\times 100.
\end{equation}

In addition to estimation accuracy, the authentication performance is evaluated from a statistical detection perspective. The probability of detection \(P_D\) and the probability of false alarm \(P_\text{FA}\) are computed under different decision thresholds, and the receiver operating characteristic (ROC) curve is plotted.

\subsection{Simulation Results}

\begin{figure}[!t]
	\centering
	\includegraphics[width=2.5in]{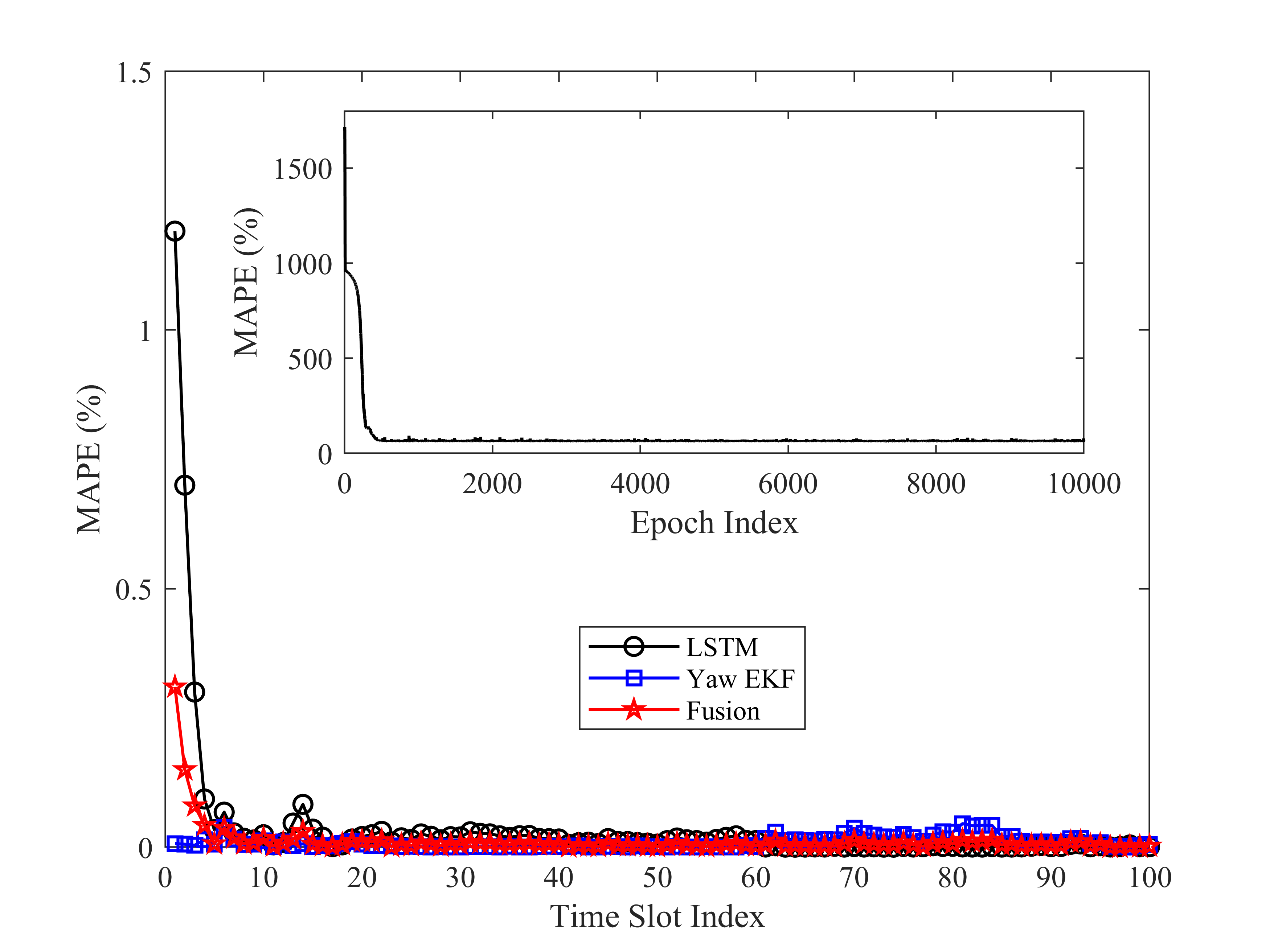}
	\caption{The convergence behavior of the proposed algorithm.}
	\label{Fig_Loss}
\end{figure}

The convergence behavior of the proposed algorithm is evaluated through simulations, as illustrated in Fig. \ref{Fig_Loss}. During the training process, the MAPE of the LSTM estimator gradually decreases and converges to a stable level. In the testing process, both the LSTM estimator and the yaw-augmented CA-EKF estimator exhibit stable MAPE performance over time. By incorporating nonlinear data-driven learning into the kinematics model-driven EKF framework via the adaptive fusion strategy, the proposed algorithm achieves a more stable and lower MAPE compared with the individual estimators, thereby demonstrating the convergence and stability of the proposed algorithm.

\begin{figure}[!t]
	\centering
	\includegraphics[width=2.5in]{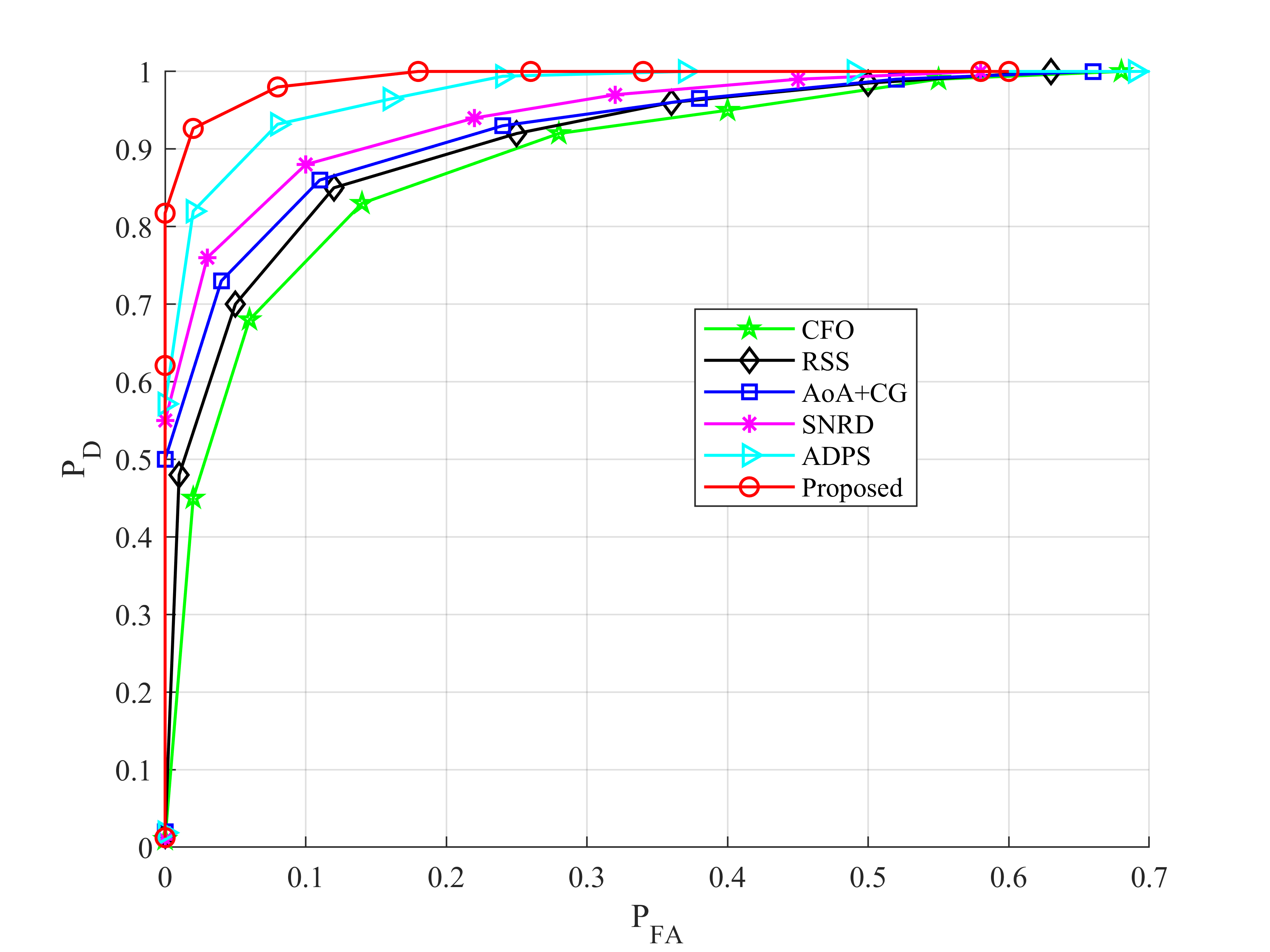}
	\caption{The ROC curves of different algorithms.}
	\label{Fig_ROC}
\end{figure}

The ROC curves of different authentication algorithms are illustrated in Fig. \ref{Fig_ROC}. It can be observed that the CFO-based authentication exhibits the worst performance, since besides the drone mobility, the wireless channel state varies across time slots even at the same location, leading to significant fluctuations in the estimated CFO. The authentication performances of the RSS-based and AOA+CG-based algorithms are comparable. While RSS mainly reflects distance-related path loss characteristics, the incorporation of AOA enables the estimation of the 3D spatial location rather than merely the distance, thereby improving the authentication accuracy. However, both the location-dependent metrics and channel-related measurements are sensitive to drone mobility and time-varying channel conditions, which may degrade the reliability of absolute feature values. In contrast, the difference of authentication metrics between adjacent time slots is more reliable than the absolute metric values, since short-term variations caused by slow channel evolution and hardware impairments tend to be partially canceled out. As a result, the SNRD-based algorithm achieves the best performance among energy-based authentication algorithms. The ADPS-based algorithm further improves the authentication performance by jointly exploiting richer signal features in the angle-delay-power domain, but its performance remains inferior to that of the proposed algorithm. This is because the proposed algorithm not only fully leverages both wireless sensing and decoded information, but also adaptively fuses the outputs of two estimators in an error-aware manner, thereby achieving enhanced robustness against nonlinear drone kinematics and time-varying channel conditions.

\begin{figure}[!t]
	\centering
	\includegraphics[width=2.5in]{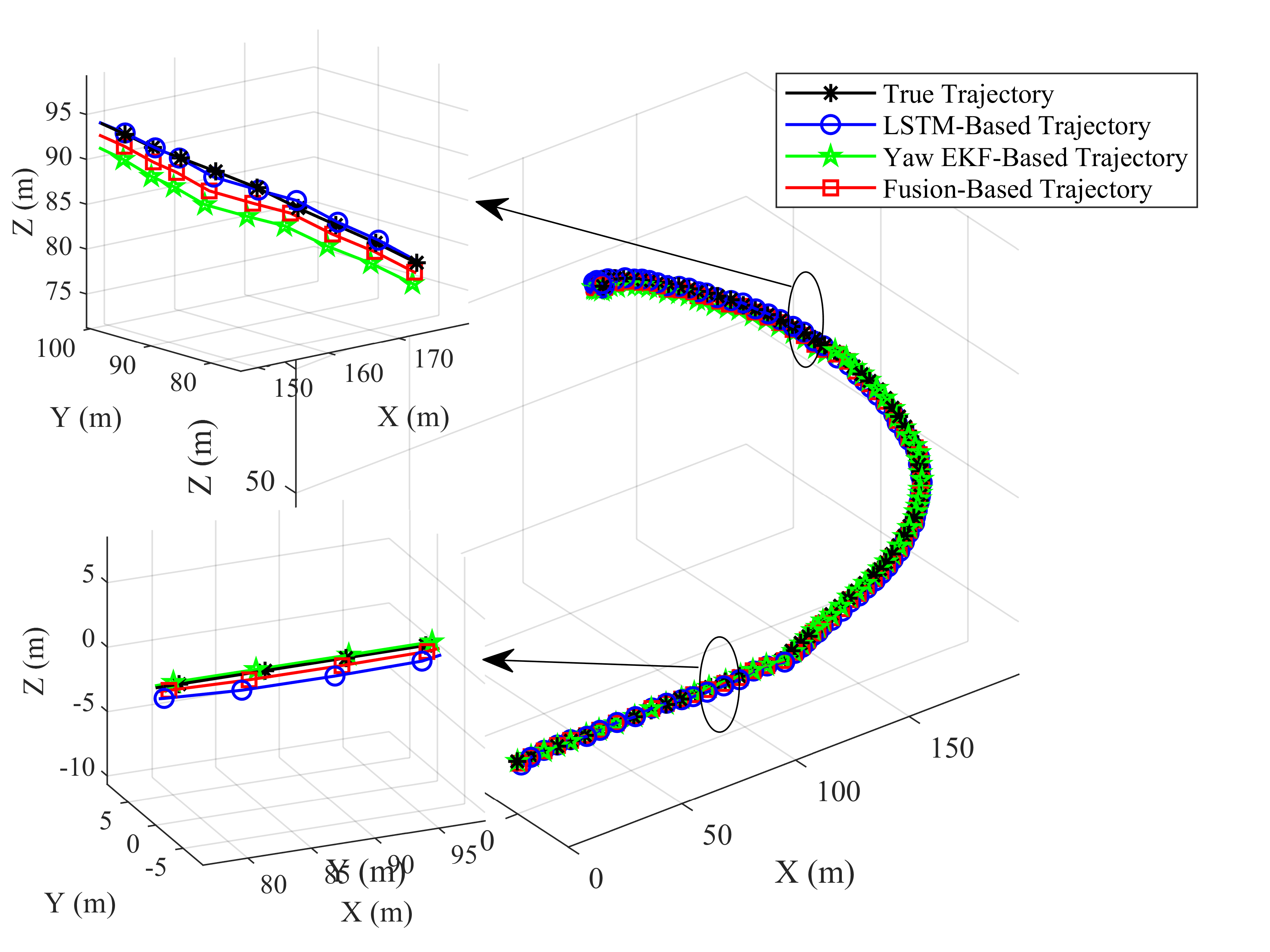}
	\caption{The estimated trajectory of the proposed algorithm under scenario 2.}
	\label{Fig_Trajectory2}
\end{figure}

To further demonstrate the effectiveness of employing two estimators and the proposed adaptive fusion strategy, ablation study results of the proposed algorithm under scenarios 2 and 3 are presented in Fig. \ref{Fig_Trajectory2} and Fig. \ref{Fig_Trajectory3}. In relatively simple motion patterns, such as straight-line trajectories, although random accelerations are present, the motion dynamics can still be well characterized by a kinematic model, and thus the yaw-augmented CA-EKF estimator yields more accurate trajectory estimates. In contrast, for complex motion patterns, such as accelerated spiral ascent or highly maneuverable flight, the LSTM-based estimator is able to better capture long-term temporal dependencies and nonlinear motion behaviors, leading to improved estimation accuracy. By adopting an error-aware adaptive fusion strategy, the LSTM outputs are utilized to compensate for and correct the EKF estimates, which enables more effective exploitation of both wireless sensing and decoded information, thereby enhancing the robustness of trajectory estimation and signal authentication.

\begin{figure}[!t]
	\centering
	\includegraphics[width=2.5in]{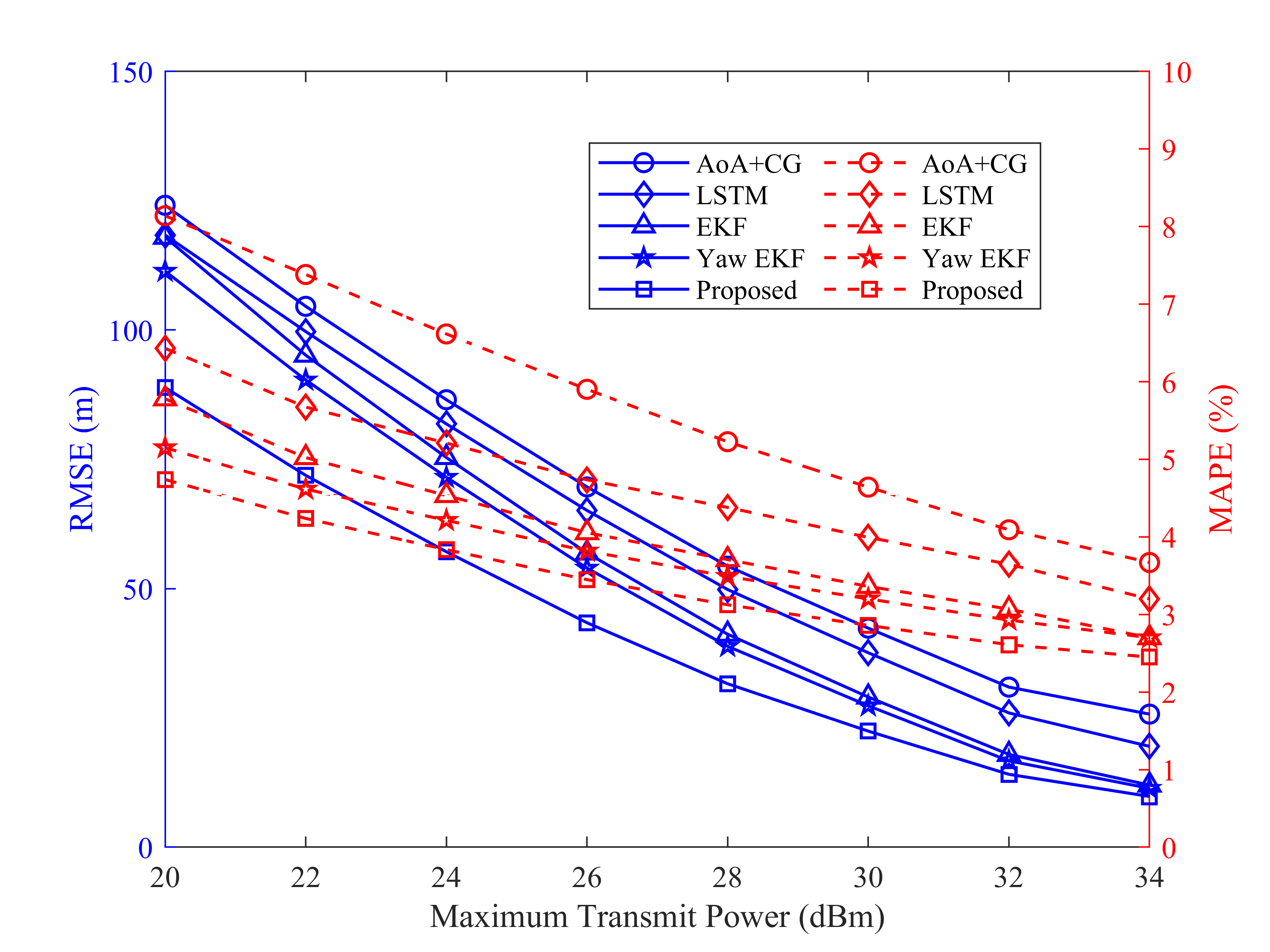}
	\caption{The trajectory estimation errors of different algorithms versus \(\tiny{P_t}\).}
	\label{Fig_Position}
\end{figure}

\textbf{\begin{figure}[!t]
		\centering
		\includegraphics[width=2.5in]{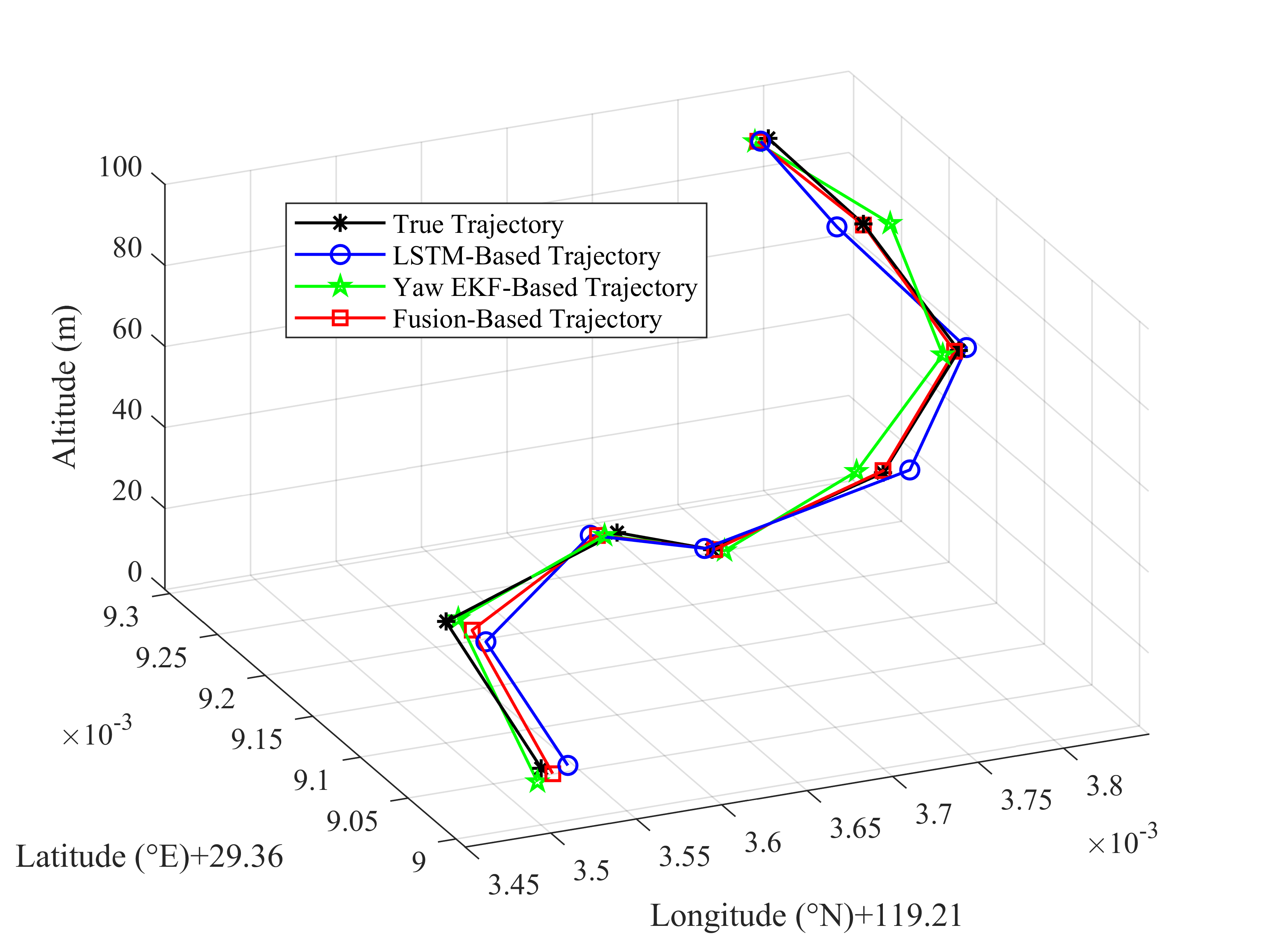}
		\caption{The estimated trajectory of the proposed algorithm under scenario 3.}
		\label{Fig_Trajectory3}
\end{figure}}
The trajectory estimation errors of different algorithms are evaluated as shown in Fig. \ref{Fig_Position}, where both the RMSE and MAPE decrease as \(P_t\) increases. Among the algorithms that perform RID frame authentication based solely on either real-time sensing or historical decoded information, the algorithm relying only on AoA and channel gain achieves the worst performance. Although the LSTM-based algorithm exploits historical decoded information, it ignores the correction gain provided by real-time sensing measurements. In contrast, the yaw augmented CA-EKF estimator makes more comprehensive use of wireless sensing information but suffers from limited robustness to time-varying channel sates and nonlinear motion dynamics, resulting in inferior estimation accuracy compared to the proposed algorithm.

The velocity estimation error of the proposed algorithm decreases as \(P_t\) increases, as illustrated in Fig. \ref{Fig_FD}. However, this improvement gradually saturates due to the presence of inherent and time-varying frequency offsets in the wireless channel, which impose a fundamental error floor on CFO-based velocity estimation.

\textbf{\begin{figure}[!t]
		\centering
		\includegraphics[width=2.5in]{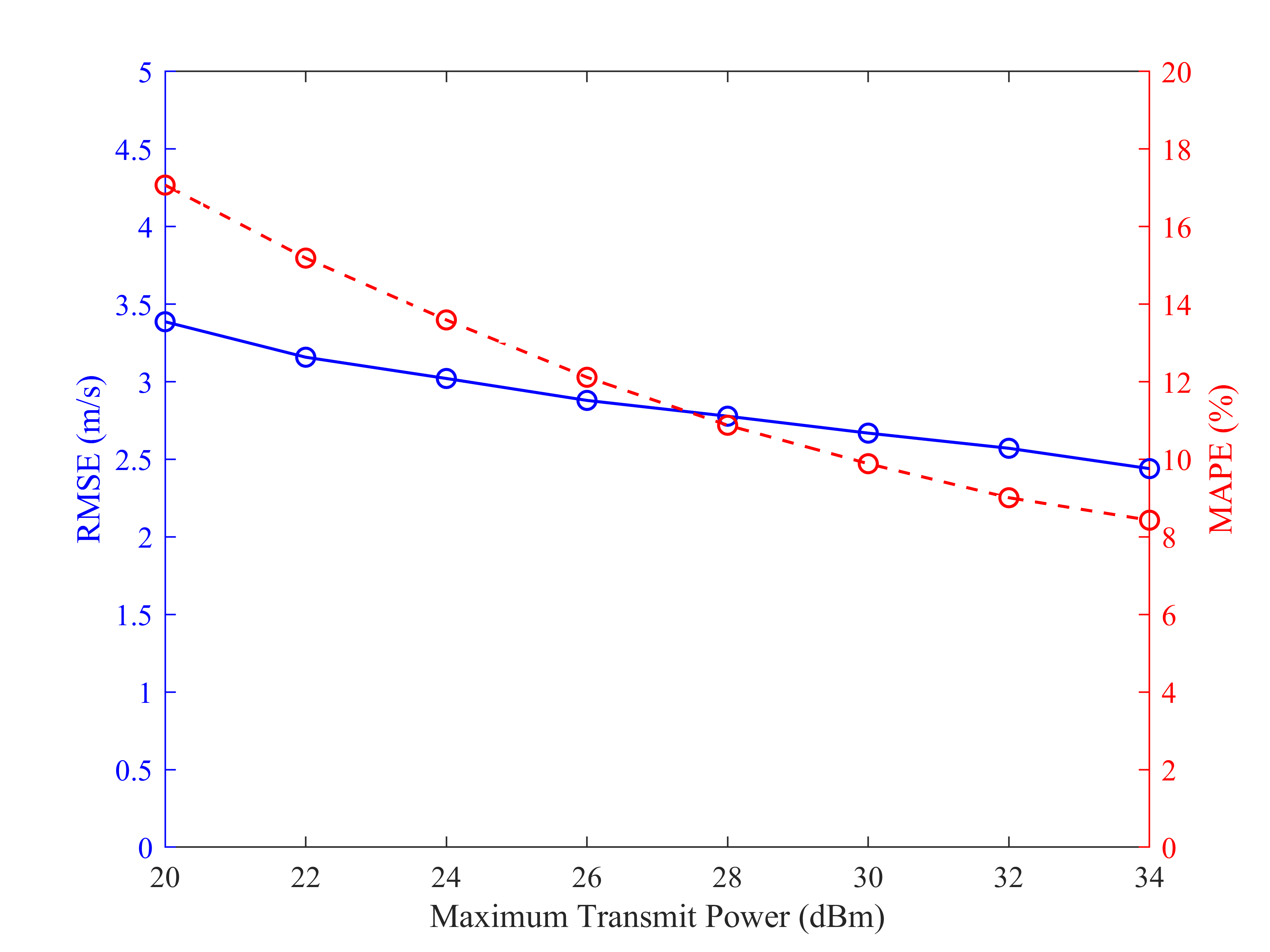}
		\caption{Velocity estimation error of the proposed algorithm versus \(\tiny{P_t}\).}
		\label{Fig_FD}
\end{figure}}
The parameter \(N_r\) affects the angular resolution of the AOA estimation and consequently influences the position estimation accuracy, as illustrated in Fig. \ref{Fig_AOA}. It can be observed that the elevation angle exhibits the largest MAPE, while the azimuth angle achieves the smallest MAPE. However, when a commonly used \(8\times8\) antenna array is employed, the performance is already close to convergence, and further increasing the array size results in significantly higher computational complexity with only marginal performance improvement.

\textbf{\begin{figure}[!t]
		\centering
		\includegraphics[width=2.5in]{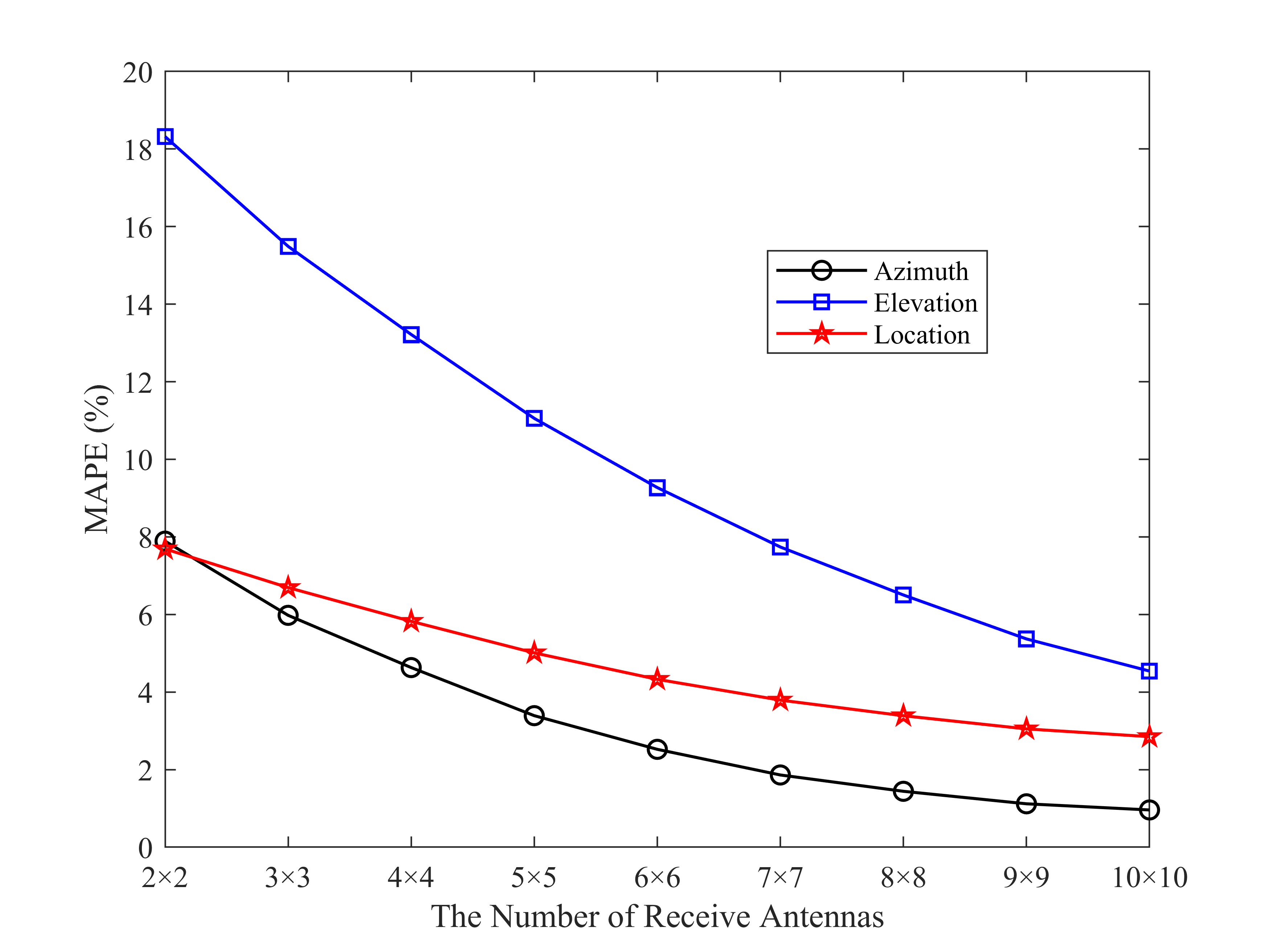}
		\caption{AoA estimation error of the proposed algorithm versus \(\tiny{N_r}\).}
		\label{Fig_AOA}
\end{figure}}
The impact of different values of \(N_r\) on the estimation performance of parameter \(N_t\) is illustrated in Fig. \ref{Fig_Ant}. It can be observed that increasing the number of \(N_r\) from \(2\times2\) to \(10\times10\) leads to a significant improvement in estimation performance. This is because, with a fixed number of transmit antennas, a larger receive antenna array increases the observation dimension, resulting in a clearer separation between the signal and noise subspaces, thereby enhancing the estimation capability of the proposed algorithm.

\textbf{\begin{figure}[!t]
		\centering
		\includegraphics[width=2.5in]{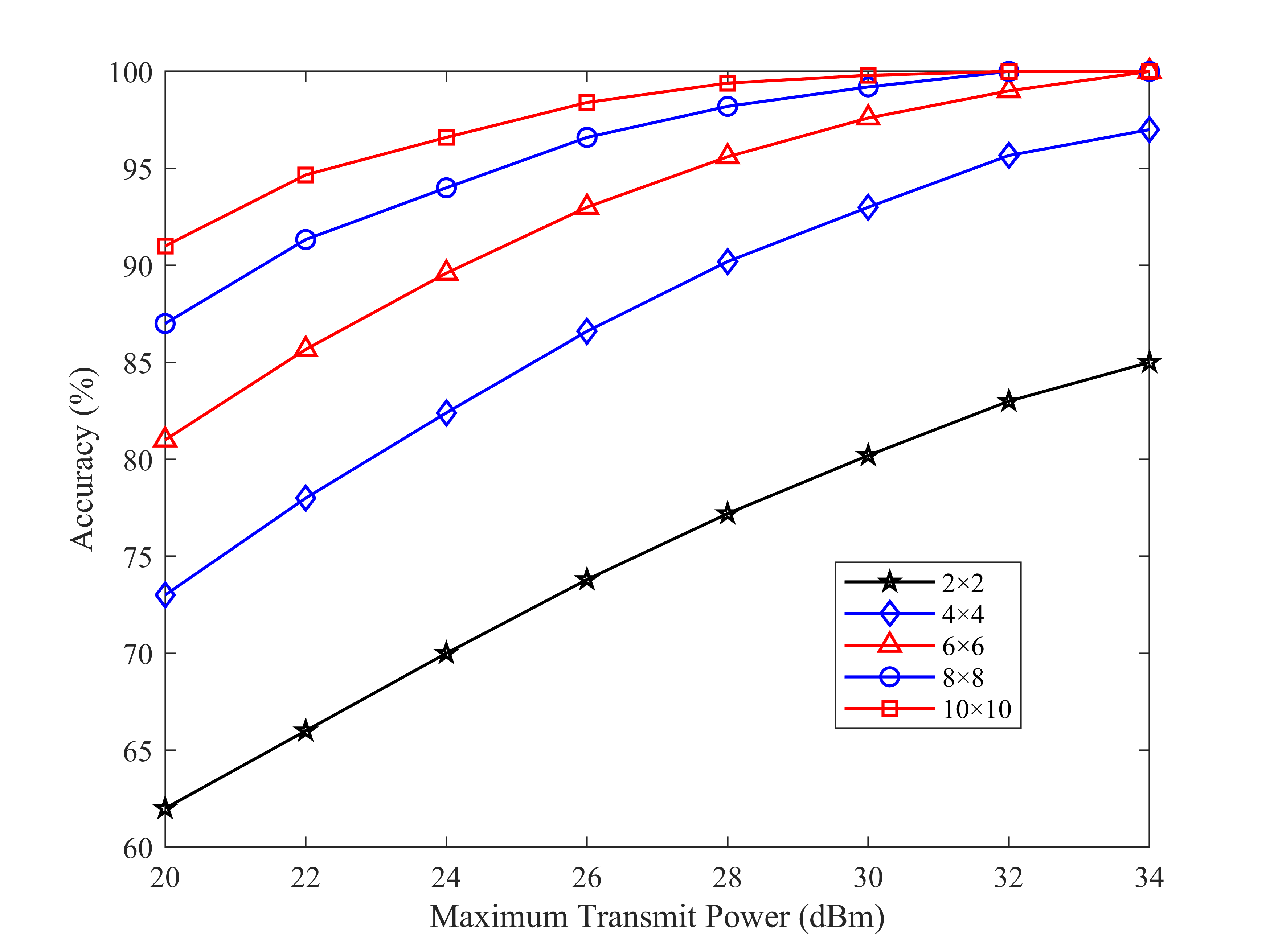}
		\caption{Estimation accuracy of \(\tiny{N_t}\) versus \(\tiny{N_r}\).}
		\label{Fig_Ant}
\end{figure}}
The parameters \(\alpha\) and \(\beta\) in \textcolor{blue}{(\ref{Eq31})} and \textcolor{blue}{(\ref{Eq36})}, which are cross-validated as shown in Fig. \ref{Fig_Para}. It can be observed that the proposed algorithm achieves the minimum MAPE when \(\alpha=0.2\) and \(\beta=0.3\). This is because such a parameter configuration enables the yaw-augmented CA-EKF to strike a proper balance between model prediction and measurement correction, allowing the observations to effectively compensate for prediction errors while avoiding excessive sensitivity to measurement noise and channel uncertainties, thereby yielding more stable and accurate state estimation.

\textbf{\begin{figure}[!t]
		\centering
		\includegraphics[width=2.5in]{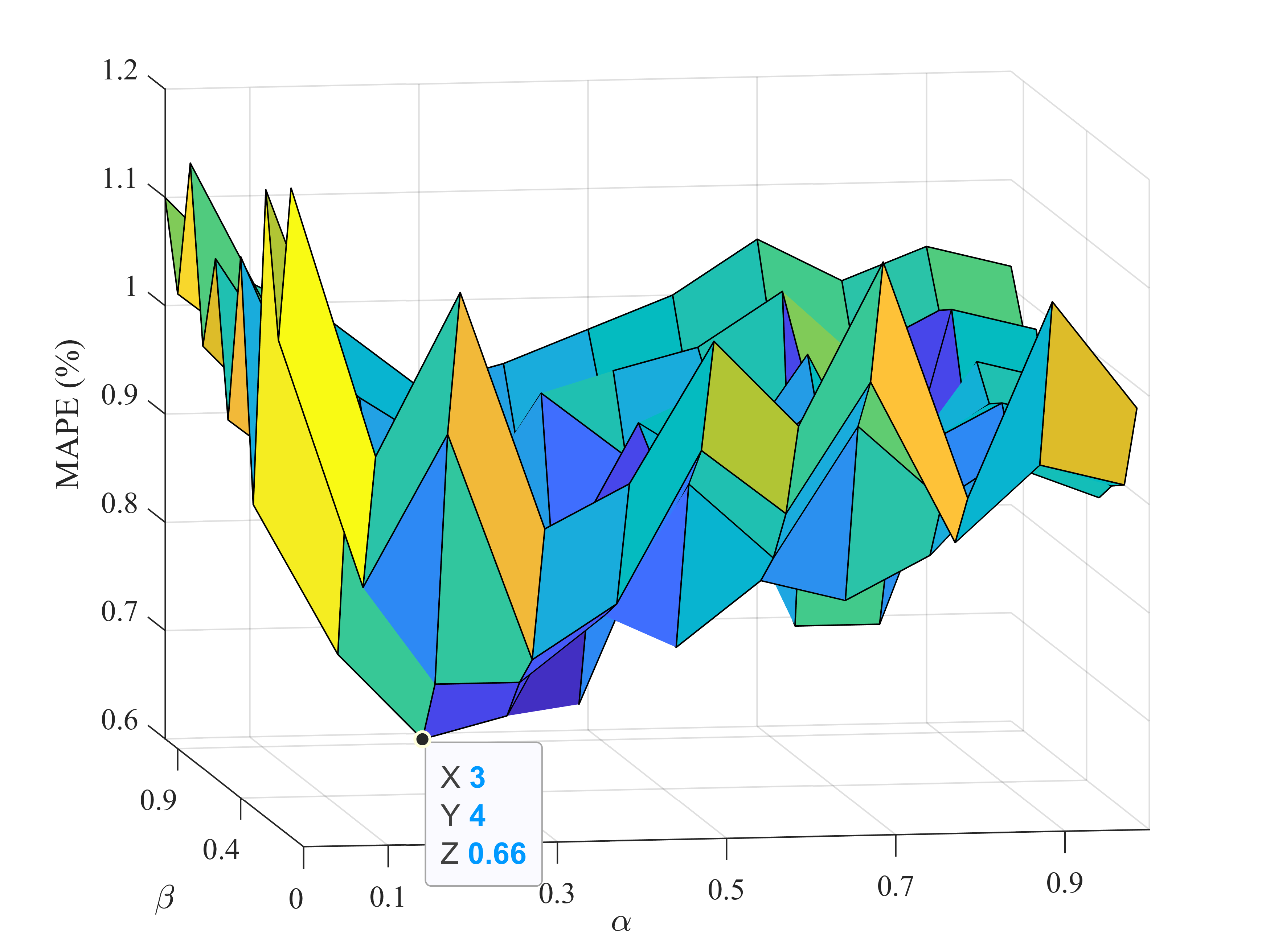}
		\caption{Location estimation error versus \(\tiny{\alpha}\) and \(\tiny{\beta}\).}
		\label{Fig_Para}
\end{figure}}

\section{Conclusion}
\label{sec5}
This paper proposed a consistency verification-based PLA algorithm for drone RID frames in realistic A2G communication environments. Motivated by the broadcast nature of RID transmissions and their lack of cryptographic protection, the proposed algorithm exploited heterogeneous sensing and decoding information inherently available in RID frames without introducing additional protocol overhead. A RID-aware sensing and decoding module was developed to extract communication-derived sensing parameters and decoding information, which were selectively employed for motion estimation and consistency verification rather than being fused into a single representation. A yaw-augmented CA-EKF was designed to incorporate real-time wireless sensing constraints and previously authenticated motion states for 3D position and motion estimation, while a data-driven LSTM-based motion estimator was introduced to handle highly maneuverable and non-stationary flight behaviors. An error-aware estimator fusion strategy was adopted to improve robustness under time-varying wireless conditions. Finally, RID frames are authenticated by jointly verifying consistency across the number of transmit antennas, motion estimation outcomes, and no-fly-zone constraints. Simulation results demonstrated that the proposed algorithm significantly improved authentication reliability and robustness compared with existing RF feature-based and motion model-based PLA algorithms under realistic wireless impairments and complex drone maneuvers.

\end{document}